\documentclass[useAMS,usenatbib]{mn2e}
\usepackage{graphicx}
\usepackage{longtable}
\usepackage{rotating}
\usepackage{natbib}
\usepackage{lscape}
\usepackage{color}
\usepackage{amsmath}	
\usepackage{amssymb}	
\usepackage{hyperref}

\renewcommand{\vec}[1]{\mathbf{#1}}
\let\oldhat\hat
\renewcommand{\hat}[1]{\oldhat{\mathbf{#1}}}

\newcommand{\mnras}{MNRAS\,}
\newcommand{\pasp}{PASP\,}

\newcommand{\aap}{A\&A\,}
\newcommand{\aaps}{A\&A\,Supp. \, \,}
\newcommand{\apj}{ApJ\,}
\newcommand{\apjl}{ApJL\,}
\newcommand{\aj}{AJ\,}

\newcommand{\nat}{Nature\,}

\newcommand{\apjs}{Astrop. J. Supp.\,}

\newcommand{\actaa}{Acta Astron.\,}

\newcommand{\na}{New Astronomy\,}

\title[]{Variability search in M\,31 using Principal Component Analysis and the {\it Hubble} Source Catalog}

\author[M.~I.~Moretti et al.]
{M.~I.~Moretti$^{1,2}$\thanks{E-mail: imoretti@oacn.inaf.it},
D.~Hatzidimitriou$^{3,1}$,
A.~Karampelas$^{1,6}$,
K.~V.~Sokolovsky$^{1,4,5}$,\newauthor
A.~Z.~Bonanos$^{1}$,
P.~Gavras$^{1}$,
M.~Yang$^{1}$
\\
$^{1}$IAASARS, National Observatory of Athens, Vas.~Pavlou \& I.~Metaxa, 15236~Penteli, Greece\\
$^{2}$INAF - Osservatorio Astronomico di Capodimonte, Via Moiariello~16, 80131~Naples, Italy\\
$^{3}$Department of Physics, National and Kapodistrian University of Athens, Panepistimiopolis, 15784~Zografos, Greece\\
$^{4}$Sternberg Astronomical Institute, Moscow State University, Universitetskii~pr.~13, 119992~Moscow, Russia\\
$^{5}$Astro Space Center of Lebedev Physical Institute, Profsoyuznaya Str.~84/32, 117997~Moscow, Russia\\
$^{6}$American Community Schools of Athens, 129 Aghias Paraskevis Ave. \&
Kazantzaki Street, Halandri, Athens GR 15234\\
}
\date{Accepted XXX. Received YYY; in original form ZZZ}

\pubyear{2018}

\begin{document}
\label{firstpage}
\pagerange{\pageref{firstpage}--\pageref{lastpage}}
\maketitle

\begin{abstract}
%+ Context
Principal Component Analysis (PCA) is being extensively used in Astronomy but not yet 
exhaustively exploited for variability search. 
%+ Aims
The aim of this work is to investigate the effectiveness of using the PCA as a
method to search for variable stars in large photometric data sets.
%+ Methods
We apply PCA to variability indices computed for light curves of 18152
stars in three fields in M\,31 extracted from the {\it Hubble} Source
Catalogue.  The projection of the data into the principal components is
used as a stellar variability detection and classification tool, capable
of distinguishing between RR~Lyrae stars, long period variables (LPVs) 
and non-variables.
%+ Results
This projection recovered more than 90\% of the known variables and revealed 38 
previously unknown variable stars (about 30\% more), all LPVs
except for one object of uncertain variability type.
%+ Conclusions
We conclude that this methodology can indeed successfully identify
candidate variable stars.
\end{abstract}


%%%%%%%%%%%%%%%%%%%%%%%%%%%%%%%%%%%%%%%%%%%%%%%%%%

%%%%%%%%%%%%%%%%% BODY OF PAPER %%%%%%%%%%%%%%%%%%

\section{Introduction}
An increasing number of time-domain surveys are producing large catalogs of multi-epoch
and multi-band photometric data, making it ever more important
to invent algorithms that efficiently detect and classify variable objects.
Multi-epoch photometry from the ground is collected by surveys for optical transients 
(e.g. Palomar Transient Factory -- \ PTF, \citealt{Law2009}; 
Catalina Real-Time Transient Survey -- CRTS, \citealt{Drake2009};
All Sky Automated Survey for SuperNovae -- ASAS-SN, \citealt{Shappee2014}),
microlensing surveys 
(MAssive Compact Halo Object -- MACHO, \citealt{Griest1991};
Optical Gravitational Lensing Experiment -- OGLE, \citealt{Udalski2008}; 
Exp\'erience pour la Recherche d'Objects Sombres -- EROS, \citealt{Kim2014}, \citealt{Tisserand2007}),  
near-infrared surveys 
(e.g. Vista Variables in the Via Lactea -- VVV, \citealt{Minniti2010}; 
The Vista near-infrared $Y$, $J$, $K_S$ survey of the Magellanic Clouds -- VMC, \citealt{Cioni2011}), 
ground-based exoplanet surveys
(Trans-Atlantic Exoplanet Survey -- TrES, \citealt{Alonso2007}; 
Super Wide Angle Search for Planets -- SuperWASP, \citealt{Butters2010}; 
Hungarian-made Automated Telescope Network -- HATNet, \citealt{Bakos2004}; 
the Kilodegree Extremely Little Telescope -- KELT, \citealt{Pepper2007})
and from space by Gaia (\citealt{Perryman2001},
\citealt{GaiaCollaboration2016}, \citealt{Clementini2016}) and Kepler (\citealt{Koch2010})
with the next generation time-domain surveys running or just around the corner 
( among them PanSTARRS, \citealt{2016arXiv161205560C};
Large Synoptic Survey Telescope -- LSST, \citealt{2008arXiv0805.2366I};
 the Next Generation Transit Survey -- NGTS, \citealt{2017arXiv171011100W}).

 Identifying variable objects in a set of light curves (LCs) is non-trivial
as the photometric measurements are often affected by correlated noise and outliers
(Sec.~\ref{sec:discussion}). Traditional variability detection techniques
include: identifying LCs with high scatter \citep[e.g.][]{Kolesnikova2008,2016MNRAS.461.3854B,2018arXiv180202303D} or the ones showing
smooth systematic variations \citep{1993AJ....105.1813W,Stetson1996,2014A&A...568A..78M},
periodicity search \citep{Fruth2012,2014ApJS..213....9D,2018AJ....155...39O,2018arXiv180201581M}
and LC template fitting \citep{1999AJ....117.1313L,2009AJ....137.4697Y,2014A&A...567A.100A,2017AJ....153..204S}.
It is desirable for a variability detection algorithm to have the
capability of detecting not only specific types of variable stars
(e.g. Cepheids) or transient events with unique signatures
(e.g. microlensing events), but also to discover all variable sources
within the capabilities of a given survey.  Once a variable object has
been detected, it is often desired to determine its variability type
from the light curve.  Methods for automatically classifying LC of
variable stars have been investigated by several authors
(\citealt{Debosscher2007}, \citealt{Paegert2014}, \citealt{Kim2016}).

The ``{\it Hubble} Catalog of Variables'' (HCV; \citealt{Gavras2017},
\citealt{Sokolovsky2017b}, \citealt{2017arXiv171111491Y}) is an ESA project aiming to develop an
algorithm for automatically detecting all variable sources within the
{\it Hubble} Source Catalog (HSC, \citealt{Whitmore2016}) which will
populate the HCV. The HSC will ultimately contain photometry of all
sources observed by the main cameras on the {\it Hubble Space Telescope
  (HST)} over its lifetime.  Despite the fact that {\it HST}
observations have been 
 extensively
used for variability studies
\citep[e.g.][]{Freedman2001,
2009AJ....137.4707M,
Clementini2009,Fiorentino2010,Fiorentino2013,DiCriscienzo2011,2013MNRAS.432.3047B,2016ApJ...830...10H}
the HSC contains a wealth of variability information that is yet to be
explored. 
 While constructing the HCV pipeline we considered 
a number of statistical characteristics, 
referred hereafter as {\it variability indices}  (Sec.~\ref{sec:varindices}), which quantify LC scatter and/or its smoothness. 
The ability of each index to identify variable sources depends on the
variability type and observing cadence. Since the {\it HST} archive
includes observations of Galactic and extragalactic fields (expected to
have stellar variable sources as well as active galaxies, supernovae
etc.) performed with various observing cadence over various time
intervals, a single variability index will not optimally find variables
in all {\it HST} fields. Here we explore a technique that may solve the
problem of selecting an optimal variability index by automatically
finding a quasi-optimal combination of multiple variability indices with
no a~priori information about the types of variable objects found in a
given field.

 \cite{Sokolovsky2017} have 
investigated the effectiveness of
applying variability indices to ground-based photometry.  The authors
proposed two robust ways of identifying variable sources in time-series
photometry of any cadence: {\it (i)}\,to use a combination of two
indices, the correlation-based inverse von~Neumann ratio and the
scatter-based Interquartile Range and {\it (ii)}\,to use the admixture
coefficient of the first principal component resulting from the principal
component analysis (PCA), as an
optimal linear combination of multiple variability indices.
 Despite ground-based data are unstable both in terms of detector 
performance as well as in PSF variability due to seeing, it was 
found that the variability-related information may be recovered
from the first significant principal component, without having to
identify which index is the most suitable for each data set in advance.

In the current study we extend the PCA-based methodology of variability
detection and apply it to variability indices derived from HSC LC of
sources in three fields in M\,31.  These fields constitute part of the
``control sample'' for the HCV, as they have been studied in terms of
variability by \cite{Brown2004} and \cite{Jeffery2011}.  Our goal is to
construct a nearly-complete list of variable sources in these fields and
evaluate the applicability of the PCA-based method to LCs
containing a much smaller number of epochs than those investigated by
\cite{Sokolovsky2017}. This technique is not specific to the HSC, but
can be applied to identify variable sources in virtually any large set
of LCs.

This paper is structured as follows: Section~\ref{sec:data} describes
the data used in this study; Section~\ref{sec:method} introduces the
PCA-based variability detection method adopted for the analysis;
Section~\ref{sec:results} presents the newly discovered variables and
compares them with previously reported variables in the studied
fields. In Section~\ref{sec:discussion} we discuss the results and in
Section~\ref{sec:conclusion}. we summarize the conclusions.

%#########################################################

\section{Data}\label{sec:data}
\subsection{The {\it Hubble} Source Catalog}
We extract the multi-epoch photometry from the HSC version~1 (HSCv1,
\citealt{Whitmore2016}), which contains 80 million detections of 30
million sources. The catalog is based on images obtained with the WFPC2,
ACS/WFC, WFC3/UVIS and WFC3/IR cameras on board the {\it HST} over a
period of 16, 15, 8 and 8 years, respectively\footnote{A second version
  of the HSC was made available during the preparation of the present
  manuscript.}.  As the {\it HST} data were collected with different
instruments, filters and observing strategies, the photometric accuracy,
data quality and cadence of observations vary greatly across the HSC,
making the detection of variable objects a non-trivial task.
The HSCv1 contains about 5 million sources with more than 2 measurements
and about 40,000 objects with more than 10 measurements obtained with
the ACS/F814W instrument/filter combination.  The F814W is available
also on the other cameras: WFPC2 and WFC3.  The F814W magnitudes of
objects listed in HSCv1 range from 15th to 26th~mag for WFPC2,
15--27\,mag for ACS and 16--28\,mag for WFC3.  The mean photometric accuracy is
  better than 0.10\,mag, while the relative accuracy is $\simeq$
  0.02\,mag at best.  HSCv1 is based on the {\it Hubble} Legacy Archive
(HLA) Data Release~8 images.

 A visit is a series of one or more exposures on a target, including
 overheads, that are executed in one or more consecutive orbits.  The
 exposures are interleaved by the time required for dithering, filter
 change, other instrumental overheads and the time the target source is
 occulted by the Earth.  All exposures taken within one visit are
 combined by the HLA into a white-light image that is used for source
 detection. For each filter used in a given visit, the images taken in
 this filter are also combined. The filter-combined images are used to
 measure brightness of the sources in each filter during the visit.
 These stacked images are used for source extraction and photometry for
 the purpose of removing image artifacts caused by cosmic ray hits.  The
 SExtractor software \citep{BertinAnouts1996} is employed to produce the
 HLA source lists providing magnitude measurements for each source.
 SExtractor is operated in dual-image mode with the white-light image
 used for source detection and filter-combined images used for
 photometry.  Since the HSCv1 photometry is visit-based, the HSCv1 light
 curves for a specific source typically contain fewer data points
 compared to the respective LC presented in the literature that are
 usually extracted from individual images within a visit.

Positions of sources detected on the visit-combined HLA images are based
on the information from {\it HST} fine guidance sensors. While the fine
guidance sensors have superb internal astrometric accuracy
\citep{2017PASP..129a2001B}, the absolute positions reported by them are
limited by the position accuracy of individual guide stars.  The
resulting absolute position errors could be as large as
$\sim$1--2\,arcsec for {\it HST} images obtained before 2005 when the
Guide Star Catalog~2 became available \citep{Lasker2008}.
 The HSC is using PanSTARRS, SDSS~\citep{2014ApJS..211...17A} and
2MASS~\citep{2006AJ....131.1163S} sources within
 the {\it HST} cameras' field of view to refine the absolute astrometry
of the HLA images.  The lists of sources detected during multiple visits
of the same sky area are cross-matched using the Bayesian technique
proposed by \cite{BudavariLubow2012}.  This results in the final
absolute astrometric accuracy of better than 0.1\,arcsec for most of the
HSC sources \citep{Whitmore2016}.

\subsection{M\,31 fields}
\label{sec:HSCv1M31}

\begin{figure} 
\begin{centering}
\includegraphics[width=0.48\textwidth,clip=true,trim=1.5cm 0.3cm 2.5cm 2.0cm]{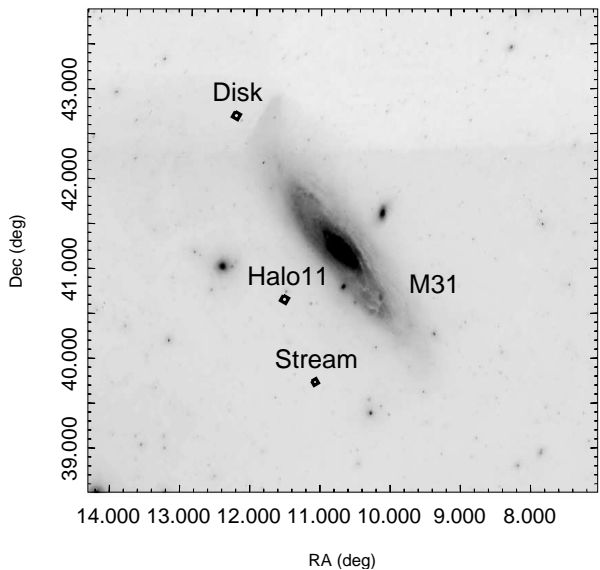} 
\caption{Finder chart of the M\,31 Halo11, Disk and Stream fields, from
  the Digitized Sky Survey. North is up, east is to the left.}
\label{fig:map}
\end{centering}
\end{figure}

We selected three fields in M\,31 observed by {\it HST}. These {\it HST}
fields were originally investigated for variability by \cite{Brown2004}
and \cite{Jeffery2011}.  The fields fall on the southeast minor axis
(``Halo11''\footnote{\cite{Brown2009} analyzed several fields along the
  south-east minor axis: one at 11, one at 21 and two at 35 kpc with
  respect to the M\,31 center; here we refer to the first one.}), on the
north-east major axis (``Disk'') and in the giant stellar stream of
metal rich stars (``Stream''; \citealt{Ibata2001};
\citealt{Kirihara2017}), as illustrated in Figure~\ref{fig:map}. The
selection of these fields was based on {\it (i)}\,the large number of
published variables (100 RR~Lyrae stars and more than 30 Long Period
Variables -- LPVs) with very good astrometry and {\it (ii)}\,the
availability of LC in two filters (F606W and F814W) in the HSCv1.
Table~\ref{tab:data} summarizes the main characteristics of these fields
and the corresponding observations: coordinates, original Program~ID,
number of HSCv1 sources, minimum, maximum and median number of {\it HST}
visits (corresponding to the number of data points in the HSCv1 LC) and
time (modified Julian date, MJD\footnote{MJD=JD-2400000.5 d}) and
magnitude range.  In the Halo11 and the Disk fields $\sim$7000 point
sources were detected in HSCv1, while $\sim$4300 sources were recorded
in the Stream field. For the Halo11 field, the observations were
obtained over a period of 40 days. The corresponding LC contain about
twice as many visits as the Stream and Disk LC, which were obtained at
lower cadence and over a shorter period of about 30 days (see discussion
in \citealt{Brown2009}).
 
\begin{table*}
 \caption{Characteristics of the M\,31 fields selected for PCA. In particular, Col. 6 and 7 indicate the number of minimum-maximum, median
 visits in F606W and F814W filters respectively.}
 \label{tab:data}
\begin{tabular}{l l@{~~~}l l r r@{~~~}r r@{~~~}r }%r@{~~~}r}
\hline
\hline
\noalign{\smallskip}
Field	& RA (deg) & Dec (deg)   & Program~ID & HSCv1    & \# Visits & \# Visits & MJD range & MJD range \\ %& F606W & F814W\\
	& J2000    & J2000       &           & sources  & F606W     & F814W     & F606W     & F814W     \\ %& (mag) & (mag)\\
\hline
Halo11 & 11.52958  & 40.71083 & GO-9453 & 7109 & 6-27, 27 & 6-33, 33 & 52610-52650 & 52611-52650        \\ %& 20.2-26.8 & 19.7-26.4\\
Disk   & 12.28583 & 42.75055  & GO-10265 & 6732 & 6-11, 11 & 6-16, 16 & 53359-53374 & 53350-53389       \\ %& 20.0-26.8 & 19.5-26.4\\
Stream & 11.07583 & 39.79222 & GO-10265 & 4311 & 6-10, 10 & 6-15, 15 & 53248-53282 & 53250-53282        \\ %& 20.2-26.8 & 19.5-26.4\\
\hline
\end{tabular}
\end{table*}

The HSCv1 sources used in our analysis satisfy the following criteria: 
{\it (i)}\,they are point sources according to the HSCv1 classification flag;  
{\it (ii)}\,they have at least 6 measurements in both the F606W and F814W LC;
{\it (iii)}\,they are located within 240\,arcsec from the center of the corresponding field;
{\it (iv)}~they have a SExtractor flag $\leq$7. Sources satisfying this
constraint may be affected by bright and nearby neighbors, bad pixels
affecting at least 10\% of the integrated area, blending with another
source, or saturated pixels. We retained data points marked as saturated
since in the HSCv1 the saturation level is set incorrectly. \footnote{We
  discarded the only source having SExtractor flag $>$7 lying in the
  Halo11 field.}  Sources within $\sim$ 0.5\,mag of the bright limit in
either filter (see Table~\ref{tab:data}) are discarded at a later stage
of the analysis to avoid false variability introduced by possible
saturation (Section~\ref{sec:results}).  Sources within 0.5\,mag of the
faint limit in either filter (Table~\ref{tab:data}) were treated with
particular care (Section~\ref{sec:results}).

\begin{figure*}
\begin{centering}
\includegraphics[width=0.48\textwidth,clip=true,trim=0cm 0cm 0cm 0cm]{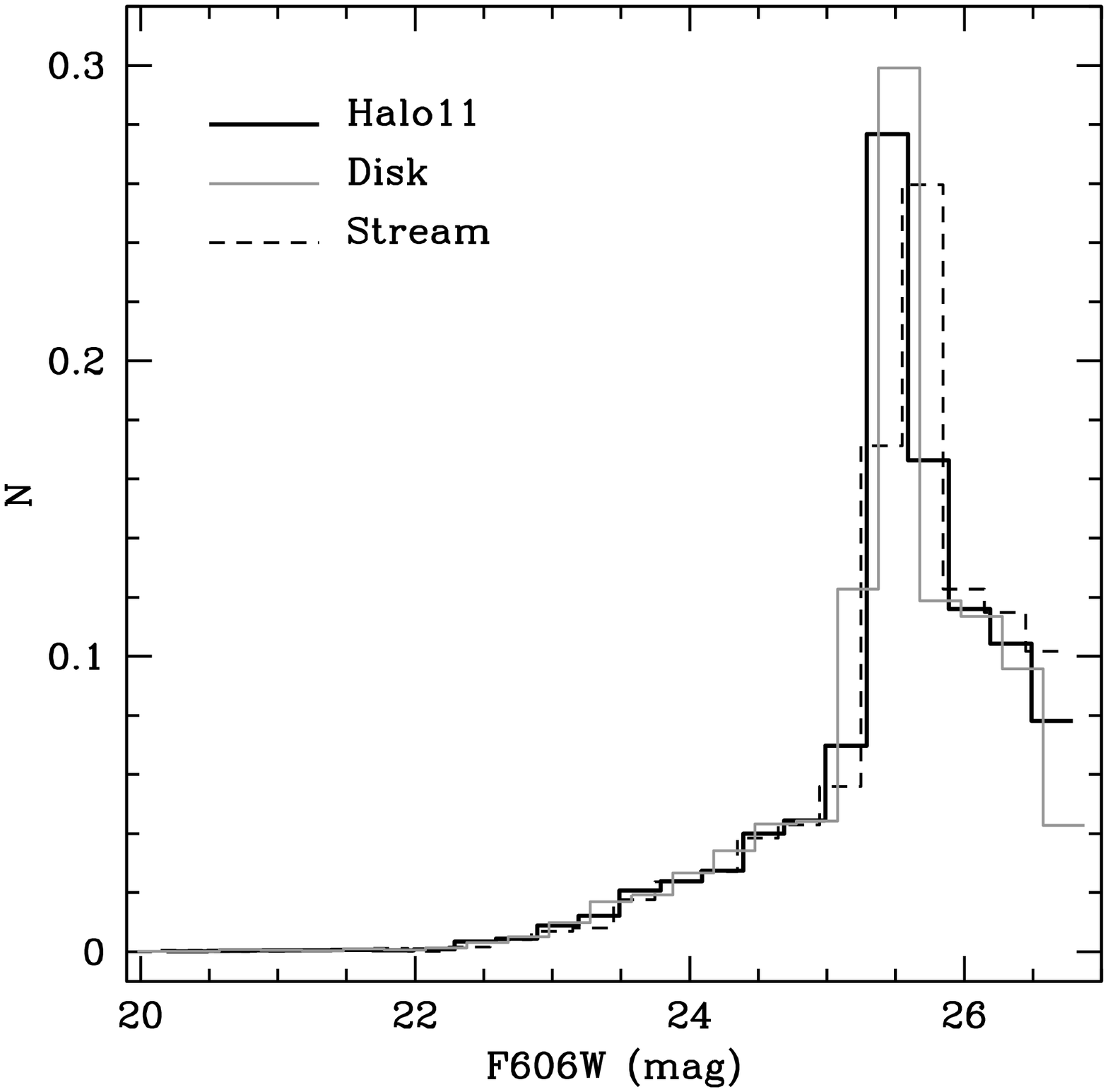}~~
\includegraphics[width=0.48\textwidth,clip=true,trim=0cm 0cm 0cm 0cm]{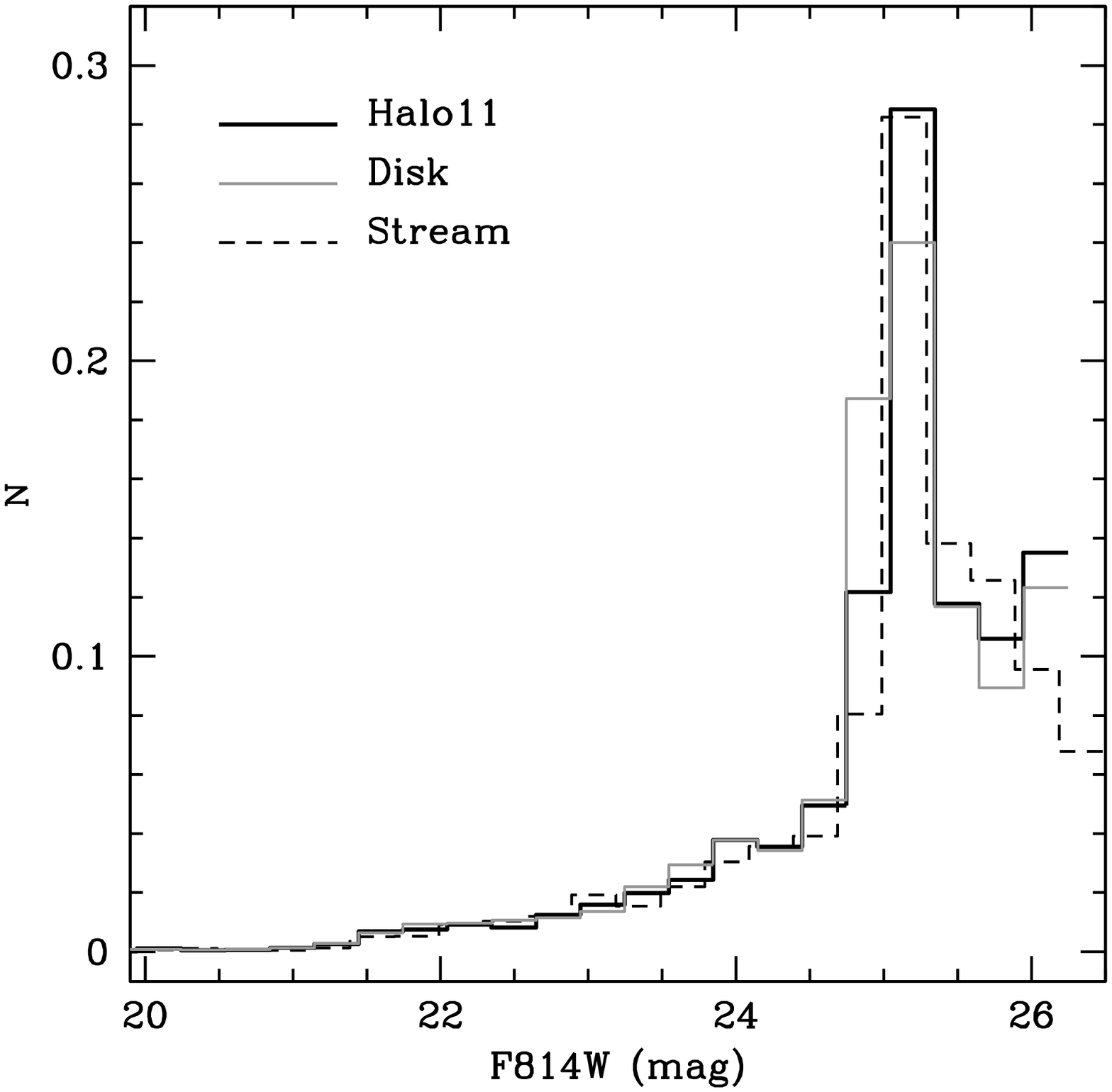}
    \caption{Normalized distribution of the mean F606W and F814W magnitudes of
 HSC sources in the three fields.}
    \label{fig:maghistall}
\end{centering}
\end{figure*}

With the above constraints, we ended up with a catalog of 18152 sources
in the three fields (``HSCv1 based sample'' hereafter).
Figure~\ref{fig:maghistall} shows the normalized magnitude histograms
for the three fields.
The histograms are similar for each filter in both bands, with a slight exception
around magnitudes 24-26 mag, possibly due to different stellar population 
properties (\citealt{Brown2006,Brown2009, Jeffery2011}; see also discussion in
Section~\ref{sec:results}).

%################################################################

\section{Methodology}
\label{sec:method}

\subsection{Variability indices}
\label{sec:varindices}
We characterize each LC by its mean magnitude and the values for 18
variability indices.  For a variable source we expect that {\it
  (i)}\,its LC has larger scatter than the LC of a non-variable object
of similar brightness (irrespective of the variability timescale) and
{\it (ii)}\,the measurements taken close in time result in similar
magnitudes (i.e. the LC is smooth) if the variability timescale is
significantly longer than the LC sampling rate. The variability indices
used here capture one or both of these characteristics. In addition to
the measured magnitudes, some indices take into account the estimated
photometric errors, as well as the order and time at which the
measurements were obtained. Table~\ref{tab:indexsummary} presents the
summary of the indices used and whether errors, order of observations or
time of observations are taken into account by the corresponding index.
A detailed discussion of these indices may be found in
\cite{Sokolovsky2017}, see also \cite{FerreiraCross2016,FerreiraCross2017}.
 While we used an HCV pipeline prototype to compute the variability
indices, all the indices listed in Table~\ref{tab:indexsummary} can also be
computed by the freely available \texttt{VaST} code \citep{2018A&C....22...28S}.

Characterizing a LC with variability indices is directly analogous to
the feature extraction  \citep{Nun2015,2016arXiv161007717C} performed 

for machine learning
classification of variable stars \citep{Debosscher2009, Richards2011, Kim2016}.  
However, here, we purposefully avoid features based on
periodicity in order to have the same set of indices characterizing
periodic, irregular and non-variable objects. Period search results may
not be reliable given the small number of observing epochs. Typically,
more than a hundred brightness measurements that randomly sample the LC
are needed for a reliable period determination
(\citealt{HorneBaliunas1986}, \citealt{Graham2013}).
 A method of characterizing LCs without defining 
variability features is proposed by \cite{2015MNRAS.451.3385K}.

Although the individual indices are useful tools for variability search, a
 combination of multiple indices 
may be an even better variability indicator, 
 as 
different indices are 
sensitive to different types of variability. 
The PCA offers a promising option 
to optimally combine variability indices, without having to decide a priori which indices are more suitable
for a given data set. 
The PCA also provides a natural 
way to combine multi-band data for variability search, as it is possible 
to add features computed from the LC in all available filters.

\begin{table}
    \caption{Variability indices used as an input for the PCA.}
    \label{tab:indexsummary}
    \begin{tabular}{r@{~~}c@{~~}c@{~~}c@{~~}c}
    \hline\hline
Index                                            & Ref. &   Errors     &    Order     &    Time     \\
    \hline
 \multicolumn{5}{c}{Scatter-based indices} \\
 \hline   
reduced $\chi^2$ statistic ($\chi_{\rm red}^2$) & (a)  & $\checkmark$ &              &              \\
weighted standard deviation ($\sigma$)          & (b)  & $\checkmark$ &              &              \\
median abs. deviation (${\rm MAD}$)             & (c)  &              &              &              \\
interquartile range (${\rm IQR}$)               & (d)  &              &              &              \\
robust median statistic (${\rm RoMS}$)          & (e)  & $\checkmark$ &              &              \\
norm. excess variance ($\sigma_{\rm NXS}^2$)    & (f)  & $\checkmark$ &              &              \\
norm. peak-to-peak amp. ($v$)                   & (g)  & $\checkmark$ &              &              \\
Stetson's~$K$ index                             & (h)  &              &              &              \\ 
\hline
 \multicolumn{5}{c}{Correlation-based indices} \\  
 \hline
Stetson's~$J$ index                             & (h)  & $\checkmark$ & $\checkmark$ & $\checkmark$ \\
weighted Stetson's~$J({\rm time})$ index        & (i)  & $\checkmark$ & $\checkmark$ & $\checkmark$ \\
clipped Stetson's~$J({\rm clip})$ index         & (d)  & $\checkmark$ & $\checkmark$ & $\checkmark$ \\
Stetson's~$L$ index                             & (h)  & $\checkmark$ & $\checkmark$ & $\checkmark$ \\
weighted Stetson's~$L({\rm time})$ index        & (i)  & $\checkmark$ & $\checkmark$ & $\checkmark$ \\
clipped Stetson's~$L({\rm clip})$ index         & (d)  & $\checkmark$ & $\checkmark$ & $\checkmark$ \\
excursions ($E_x$)                              & (j)  & $\checkmark$ & $\checkmark$ & $\checkmark$ \\
autocorrelation ($l_1$)                         & (k)  &              & $\checkmark$ &              \\
inv. von~Neumann ratio ($1/\eta$)               & (l)  &              & $\checkmark$ &              \\
$S_B$ statistic                                 & (m)  & $\checkmark$ & $\checkmark$ &              \\
    \hline
    \end{tabular}
\begin{flushleft}
References: 
(a)~\cite{deDiego2010}, 
(b)~\cite{2018arXiv180202303D}, 
(c)~\cite{Zhang2016}, 
(d)~\cite{Sokolovsky2017}, 
(e)~\cite{Rose&Hintz2007}, 
(f)~\cite{Nandra1997}, 
(g)~\cite{Brown1989}, 
(h)~\cite{Stetson1996}, 
(i)~\cite{Fruth2012}, 
(j)~\cite{Parks2014}, 
(k)~\cite{Kim2011}, 
(l)~\cite{2009MNRAS.400.1897S}, 
(m)~\cite{Figuera2013}. 
\end{flushleft}
\end{table}

\subsection{Principal Component Analysis}
 The PCA \citep{Pearson1901} is extensively used in Astronomy 
(e.g. \citealt{Bailer1998}, \citealt{Fiorentin2007}, 
\citealt{Yip2004}, \citealt{Karampelas2012}, \citealt{Steiner2009}).
It has also been employed for variable star detection using
multi-band LCs \citep{Suveges2012, Sokolovsky2017}, as first suggested by \cite{Eyer2006}.
 PCA linearly and orthogonally 
transforms a data set of $m$ quantities (where each data point is
represented by a vector, $\vec{x_{j}}$, in the $m$-dimensional space) onto a new set of $m$ uncorrelated axes 
(the eigenvectors of the variance-covariance matrix of the data), where the data variance is being
emphasized. These eigenvectors are called the principal components (PCs). 
Each observation $\vec{x_{j}}$ of the original data is expressed as 
\begin{equation}
\label{eq:pca}
\vec{x_{j}} = \sum_{i=1}^{m} a_{j,i} \cdot \vec{PC_{i}}
\end{equation}
 where $a_{i}$ is the admixture coefficient of the principal component
$\vec{PC_{i}}$. The coefficients $a_{i}$ are the coordinates of the data point $\vec{x_{j}}$
in the new axes. 

PCs are ordered so that $\vec{PC_{1}}$ accounts for the highest data variance, 
$\vec{PC_{2}}$ (uncorrelated to $\vec{PC_{1}}$) incorporates most of the
remaining variance and so on. Thus, along the sequence from
$\vec{PC_{1}}$ to
$\vec{PC_{m}}$, information, assumed to be represented by the data variance,
extends from widespread, to rare, to noise.  The amount of the variance
corresponding to each PC is the respective eigenvalue $\lambda_{i}$
divided by the sum of all the eigenvalues. This relation originates from
the diagonalization of the data variance-covariance matrix $C$, achieved
through the PCA implementation.  Each element $c_{ij}$ of $C$ is the
covariance between the $i$-th and the $j$-th variables. If $i=j$, then
$c_{ii}$ is the variance of the $i$-th variable.  The
variance-covariance matrix $C^{'}$ of the transformed data is diagonal,
having $c_{ij}^{'}=0$ and $c_{ii}^{'}=\lambda_{i}$. The larger the
$\lambda_{i}$, the higher the variance represented by $\vec{PC_{i}}$.

Since the lower-order PCs do not represent any useful information, each
observation can be 
 approximately expressed by replacing $m$ with $r$ in Eq.~(\ref{eq:pca}) 
 where $r<m$. 
There is no
standard procedure for deciding how many principal components should be
kept.  If $r=2$ or $r=3$, the data can be visualized in two or
three dimensions, using the corresponding admixture
coefficients $a_{1}$, $a_{2}$ and $a_{3}$. The resulting dimensionality
reduction/data compression is among the advantages of the PCA. 
 The PCA is 
an unsupervised (no need for training data) and non-parametric (no need for tuning)
procedure that provides a linear decomposition. 
On the other hand, PCA may not perform very well when the
processes dominating the original data are not linear, although
alternatives do exist such as kernel-PCA (e.g. \citealt{Ishida2013}) and
robust-PCA (e.g. \citealt{deSouza2014}).  PCA is also data-dependent,
thus requiring a standardization of the original data, if the data
variables express different characteristics
 (Sec.~\ref{sec:pcainm31fields}).

\subsection{PCA implemented in the M\,31 fields}
\label{sec:pcainm31fields}
The input characteristics for the PCA were the 18 variability indices
listed in Table~\ref{tab:indexsummary} and the mean magnitude computed
for the LCs in each filter (F606W and F814W), resulting in a total of 38
parameters, for all sources detected in the three selected M\,31 fields
(Halo11, Disk and Stream).  The variability indices were first
standardized to zero average and unit variance.  Standardization is
necessary since the variance-enhancing PCA is data dependent and the
input characteristics express different quantities.  High values of an
index may result in numerically large variance, although this index may
not contain useful variability information. The standardized index
values express the deviation from the average in multiples of the
standard deviation, making the various indices comparable. By
construction, large values of the variability indices (subsequently,
positive values well above the zero average of the standardized indices)
are expected to correspond to variable stars.

\begin{figure}
\begin{centering}
\includegraphics[width=0.48\textwidth]{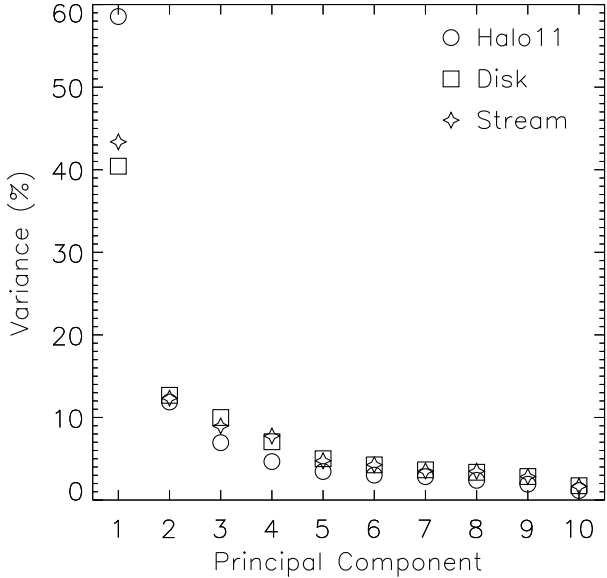}
    \caption{Scree plot representing the variances of the 10 most 
    significant Principal Components.} 
    \label{fig:scree}
\end{centering}
\end{figure}

The scree plot presented in Figure~\ref{fig:scree} illustrates the
percentage of the total variance that corresponds to 
the 10 most significant PCs. 

High variance is caused by high values of the
variability indices of variable sources, thus linking the most
significant principal components to variable sources. 
 About 60\% 
of the data variance is captured by the first two 
PCs
(average value  
for the three fields).

\subsection{Selection of candidate variables}
\label{sec:selection}

The detectability of variable sources hinges on the following
assumptions: {\it (a)} The sample contains variables and therefore the
variability indices by construction contain variability information from
the LC. {\it (b)} Variability is mainly encoded in $\vec{PC_{1}}$ and
$\vec{PC_{2}}$ and their respective values (admixture coefficients $a_1$ and
$a_2$.)  The high values of the variability indices of the variable
stars and their respective high variance ``feed'' the most significant
principal components, since PCA preferentially highlights
variance. Therefore, the variability information is expected to be
encoded in the most significant principal components. {\it (c)}
Variable stars are found in sparse regions on the $a_1$--$a_2$ plane,
while constant stars have $a_1$ and $a_2$ values close to
zero. Additionally, the observational fact that the majority of stars
appear to be constant  (at the level of photometric accuracy expected
from HSC), ensures that they will form the distinct dense area
in the $a_1$--$a_2$ plane, making the variables more easily
distinguishable. Thus, sparseness in the $a_1$--$a_2$ plane emerges from
variable stars being rare and having wider $a_1$ and $a_2$ ranges
compared to the constant stars.
 
\begin{figure*}
	\includegraphics[width=1.0\columnwidth]{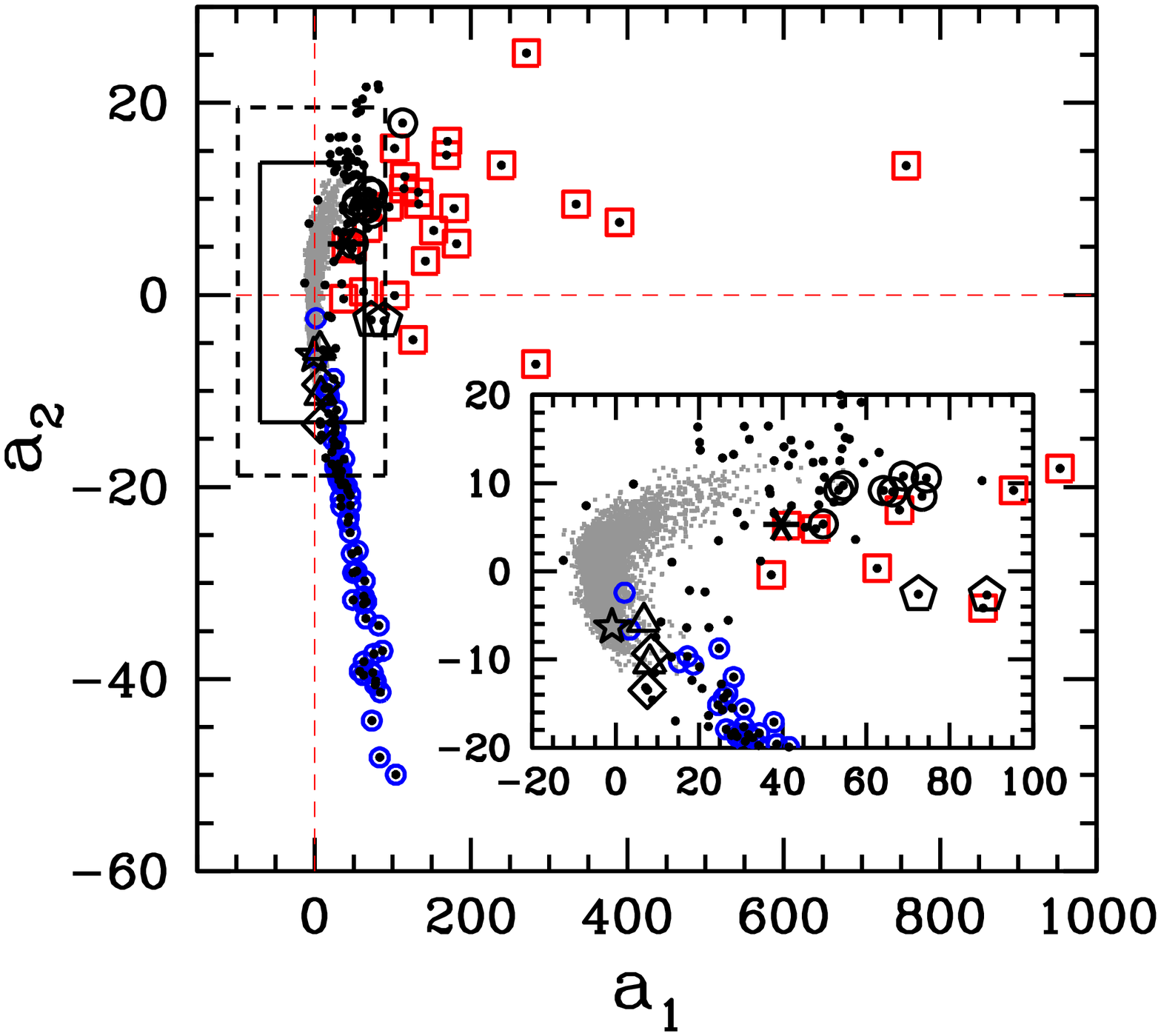}~~~~
	\includegraphics[width=1.0\columnwidth]{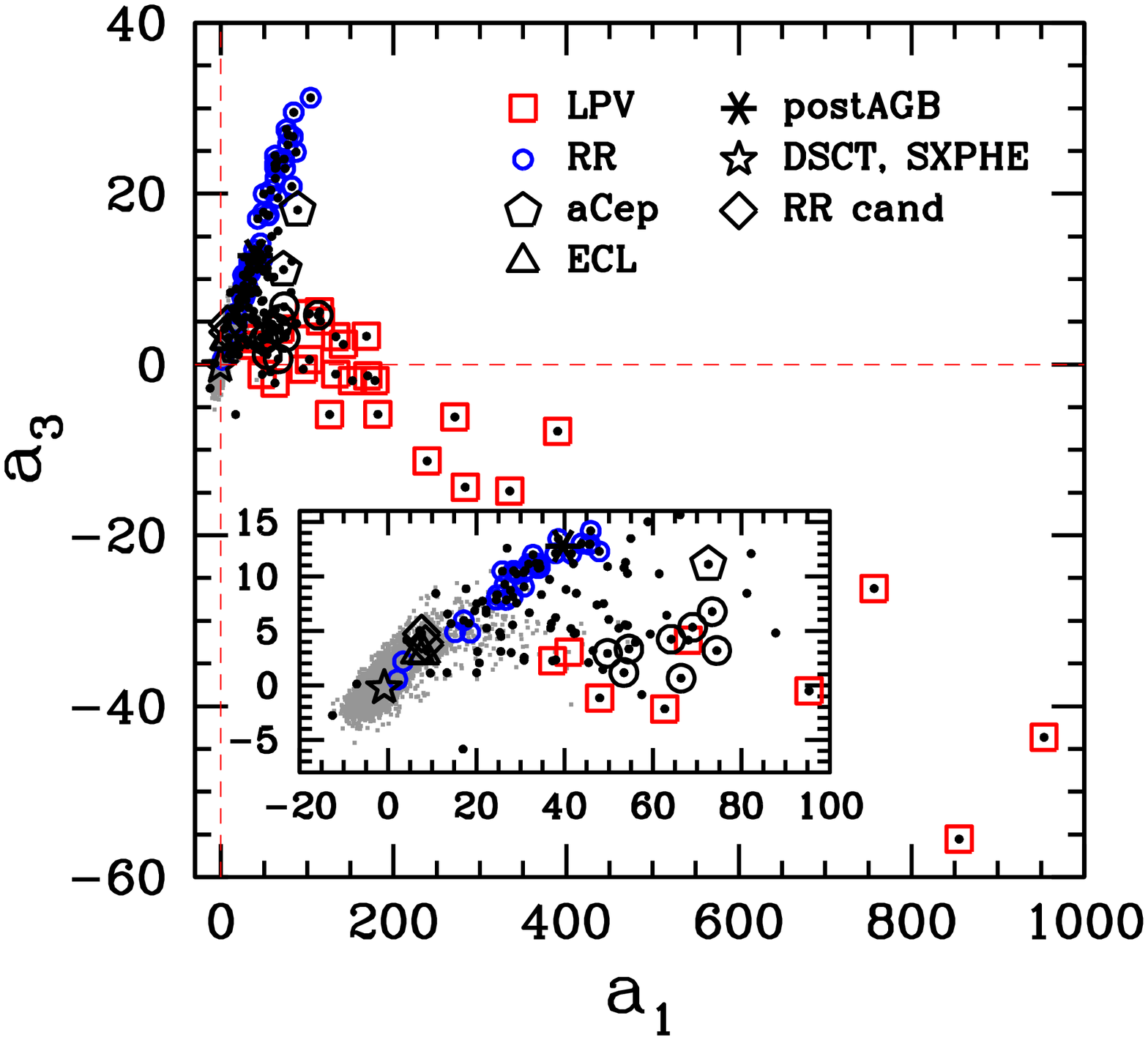}
        \caption{The location of candidate variables (black points) on
          the 
admixture coefficient planes 
for 
M\,31 Halo11. 
The known LPVs 
          (red open squares), 
RR~Lyrae stars (small blue open circles),
          RR~Lyrae star candidates (black open diamonds), eclipsing
          binaries (black open triangles), the dwarf Cepheid (black open
          star), the anomalous Cepheids (black open pentagons), the
          post-AGB star (black asterisk) are labeled, along with all
          the sources (grey dots). The newly discovered variables are
          marked with big black open circles. The solid box corresponds
          to the initial step of the selection process (step {\it (i)} in Sec.~\ref{sec:selection}), while the dashed
          box indicates the region where the second step ({\it (ii)} in Sec.~\ref{sec:selection}) of the process
          was applied. Insets present the most crowded areas.}
    \label{fig:halo_admixture}
\end{figure*}

 As the first two PCs summarize most of the data variance, 
we exploit the corresponding admixture coefficients $a_1$ and $a_2$ to
identify variable sources.
It is noted that the PCs are global for each
field (all stars share the same $\vec{PC_{i}}$), while the admixture
coefficients are star specific (each star has its own set of $a_i$).
The upper left panel of Figure~\ref{fig:halo_admixture} shows the
distribution of all sources in the Halo11 field on the $a_1$--$a_2$
plane. The great majority of sources occupy a distinct dense region
around zero $a_1$ and $a_2$ values and are interpreted as ``constant''
sources.  A small number of sources, with high absolute values of $a_1$
and $a_2$, lie outside this dense locus and are expected to correspond
to variable sources.  Similar plots for the Stream and Disk fields can
be found in the Appendix (top left panel of
Figures~\ref{fig:disk_admixture}, ~\ref{fig:stream_admixture}).
We employ the following two-step approach to select candidate
variables.

{\it (i)}\,{\it Initial selection of candidate variables}\\ Any star
with $a_1$ or $a_2$ values that differs at least three standard
deviations ($3\sigma$) from the corresponding median values (i.e.\ outside the black
solid line box in the upper-left panel of
Figure~\ref{fig:halo_admixture}), was considered to be a variable star
candidate. The $a_1$ and $a_2$ medians, dominated by the low $a_1$ and
$a_2$ values of the constant stars, would be low as well. The standard
deviations, dominated by the extreme $a_1$ and $a_2$ values of some
variable stars, will be relatively high.  Thus, this criterion is
expected to ensure detection of actual variable stars.

{\it (ii)}\,{\it Selection fine-tuning}\\ Even inside the solid line box
of Figure~\ref{fig:halo_admixture} there are sources near but not inside
the dense zone of, presumably, constant stars (that is not rectangular
in shape).  To retrieve these candidates we used the average (Euclidean)
$a_1$--$a_2$ distance of each source inside the solid line box to its
three nearest neighbors (hereafter d$_k$ for the $k$-th star), the
median source-to-source distance inside the box ($D_{inside}$), and the
median source-to-source distance outside the box ($D_{outside}$), in an
equal area lying between the solid and the dashed lines of
Figure~\ref{fig:halo_admixture}.

By construction, $D_{inside}$ characterizes constant stars, as the vast
majority of sources inside the box are expected to be non-variables. On
the other hand, most of the stars contained in the outer box are
expected to be variables. Thus, for the $k$-th inner box star to be a
variable candidate, its distance $d_k$ is expected to have a value
closer to $D_{outside}$ rather than to $D_{inside}$, that is $|d_k -
D_{outside}| < |d_k - D_{inside}|$.

The resulting candidate variables are marked with black dots on the
$a_1$--$a_2$ selection plane for M~31 Halo11 (upper
left panel of Figure~\ref{fig:halo_admixture}).
These candidate variables are validated and compared against published
variables in Section~\ref{sec:results}.  The candidate variables are
also marked on higher order admixture coefficient planes (remaining
panels of ~Figure \ref{fig:halo_admixture}) 
 which are further discussed in Section~\ref{sec:physinterpretationofpc}.
Similar plots are available for the Disk and Stream fields in the Appendix
(Figures~\ref{fig:disk_admixture}, ~\ref{fig:stream_admixture}).

\section{Results}
\label{sec:results}
\begin{figure}
	\includegraphics[width=1.05\columnwidth]{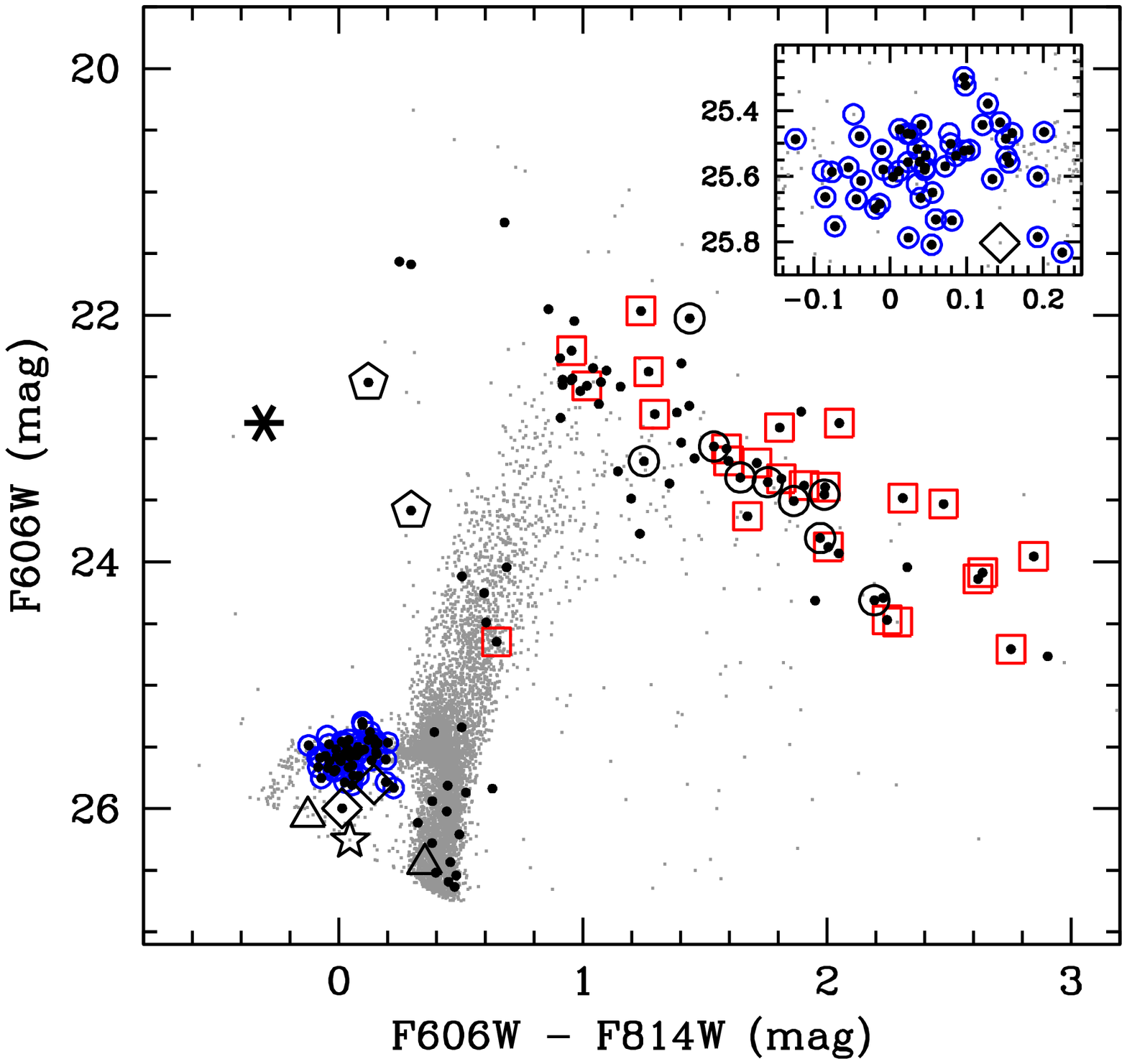}
    \caption{Color-magnitude distribution of sources in the Halo11
      field. All sources are represented with grey dots, the PCA
      variable candidates are shown with black points. The new and known
      variables are labeled as in Figure~\ref{fig:halo_admixture}. The
      inset presents the HB region.}
    \label{fig:cmdHalo11} 
\end{figure}

\begin{figure}
	\includegraphics[width=1.05\columnwidth]{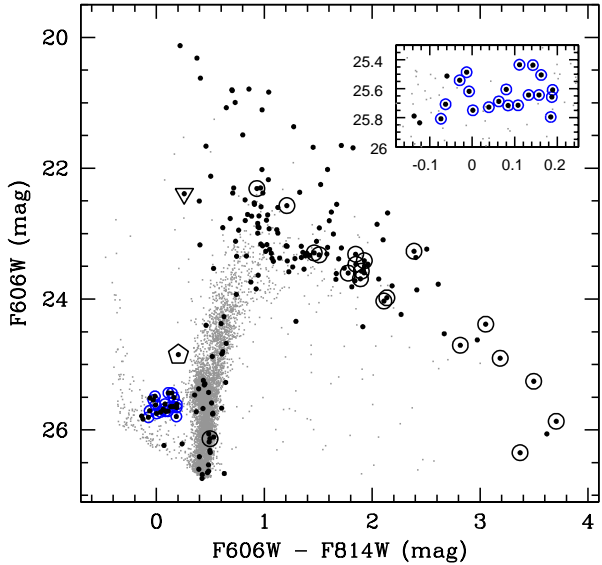}
    \caption{Same as Fig.~\ref{fig:cmdHalo11}, but for the Disk
      field. The literature variables are from 
\protect\cite{Jeffery2011};
      the classical Cepheid is marked by a reversed open triangle.}
    \label{fig:cmdDisk} 
\end{figure}

\begin{figure}
	\includegraphics[width=1.05\columnwidth]{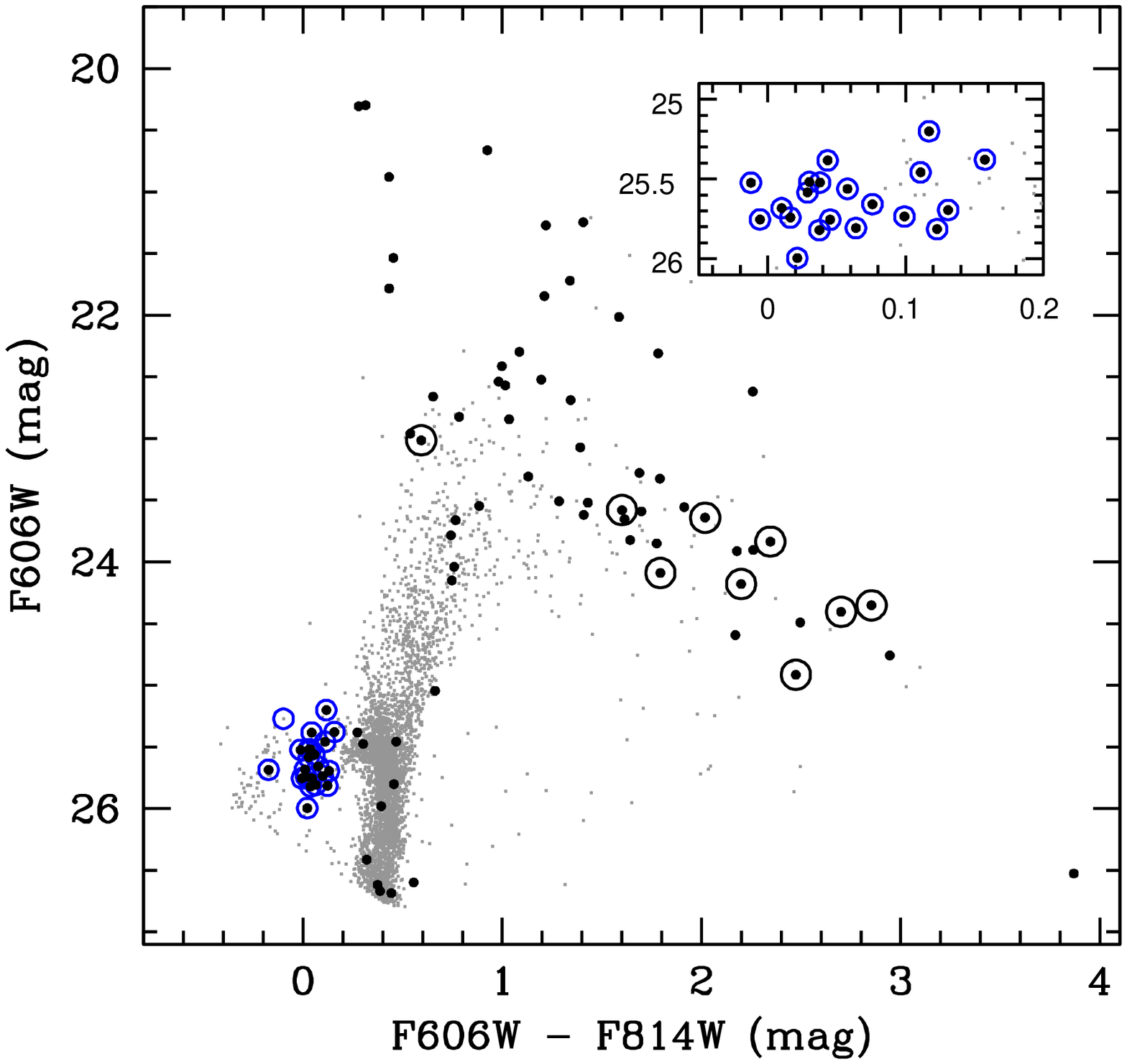}
    \caption{Same as Fig.~\ref{fig:cmdHalo11}, but for the Stream
      field. The literature variables are from 
\protect\cite{Jeffery2011}.
}
    \label{fig:cmdStream} 
\end{figure}

The two step selection procedure described in the previous section,
resulted in a total of 156 candidate variables in the Halo11 field
(2.2\% of the total number of sources), 192 in the Disk field (2.9\%)
and 88 in the Stream field (2.0\%). 
Figures~\ref{fig:cmdHalo11}, \ref{fig:cmdDisk},
\ref{fig:cmdStream}  show the color
magnitude diagrams (CMD) for the three fields, respectively. The sources
lie on the red giant (RGB), subgiant (SGB), asymptotic giant (AGB) and
horizontal branches (HB). In most cases, main sequence stars lie below
the detection limit, with the exception of the Disk field where they are
detected at $\sim$23rd magnitude, indicating the presence of a younger
population (see also \citealt{Jeffery2011}, \citealt{Brown2006,Brown2009}).
The location of the candidates on the CMD is also shown (black points in
Figures~\ref{fig:cmdHalo11},~\ref{fig:cmdDisk},~\ref{fig:cmdStream}).  
Most of the candidate variables are associated with the
HB where one expects to find RR~Lyrae stars, with the AGB where one
would expect to find LPVs and with the supergiant region. Several
candidates lie on the RGB and SGB and might be eclipsing binaries.  We
consider as reliable variable candidates stars fainter than 21.0\,mag in
F606W (20.5 in F814W), corresponding to the saturation limit derived
from the raw {\it HST} images.

\subsection{Recovery of known variables}
\label{sec:BrownComparison}

To compare the list of variable stars published by \cite{Brown2004} and
\cite{Jeffery2011}, to our list of candidate variables
(Section~\ref{sec:selection}) and evaluate the effectiveness of our
variability selection technique, we first need to ascertain whether a
published variable is present in the HSCv1 based sample, as defined in
\ref{sec:HSCv1M31}.  Some of the published variables may be excluded
due to differences in image processing and source extraction strategies
used to create the HSC and the ones employed by \cite{Brown2004} and
\cite{Jeffery2011}. For the same reason some of the sources that could
not be used in the original studies may become useful for variability
search through the HSC.

 Most of the published variables have an HSCv1 counterpart within about
 0.6\,arcsec, indicating that the published and the HSCv1 astrometry are
 in good agreement.  For RR~Lyrae variables the validity of HSCv1
 counterparts to the published variables was confirmed by visual
 inspection of the LC folded with the published period.  In the case of
 LPVs published by \cite{Brown2004}, no time series data were provided
 by the authors, thus the matching was performed based only on the
 positional coincidence and visual comparison of HSCv1 LCs with the LC
 plots from the paper.  Table~\ref{tab:literature} summarizes the
 comparison results listing the number of published variables, the
 number of published variables with a counterpart within the HSCv1 and
 the number of published variables recovered in this work.
Figures~\ref{fig:halo_admixture}, ~\ref{fig:cmdHalo11}, ~\ref{fig:cmdDisk}, ~\ref{fig:cmdStream}
also show the known variables highlighted with different symbols as indicated in the legend of 
the upper right panel of Figure~\ref{fig:halo_admixture}.

\begin{figure}
\begin{centering}
\includegraphics[width=0.48\textwidth]{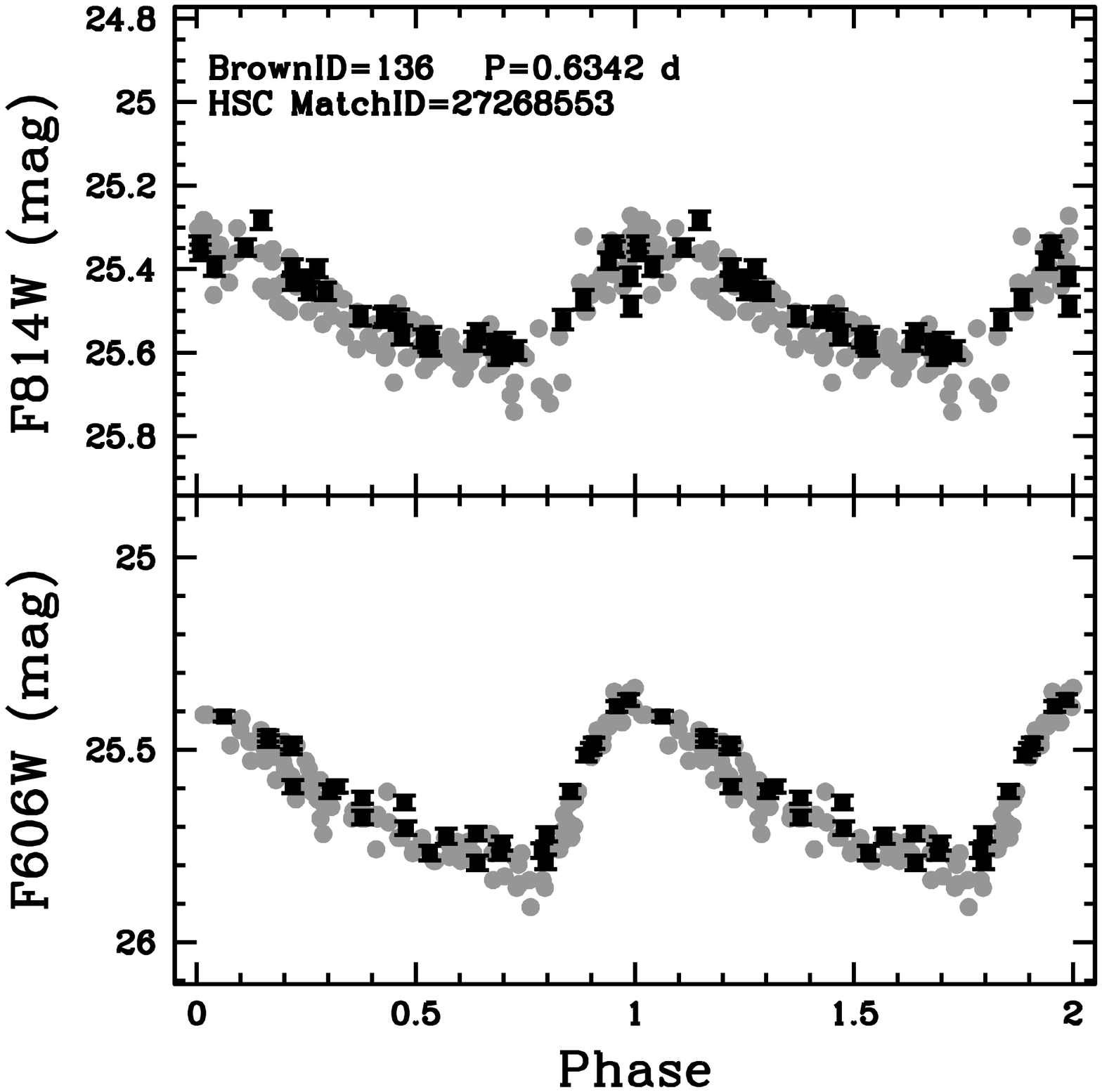}\\ 
\includegraphics[width=0.48\textwidth]{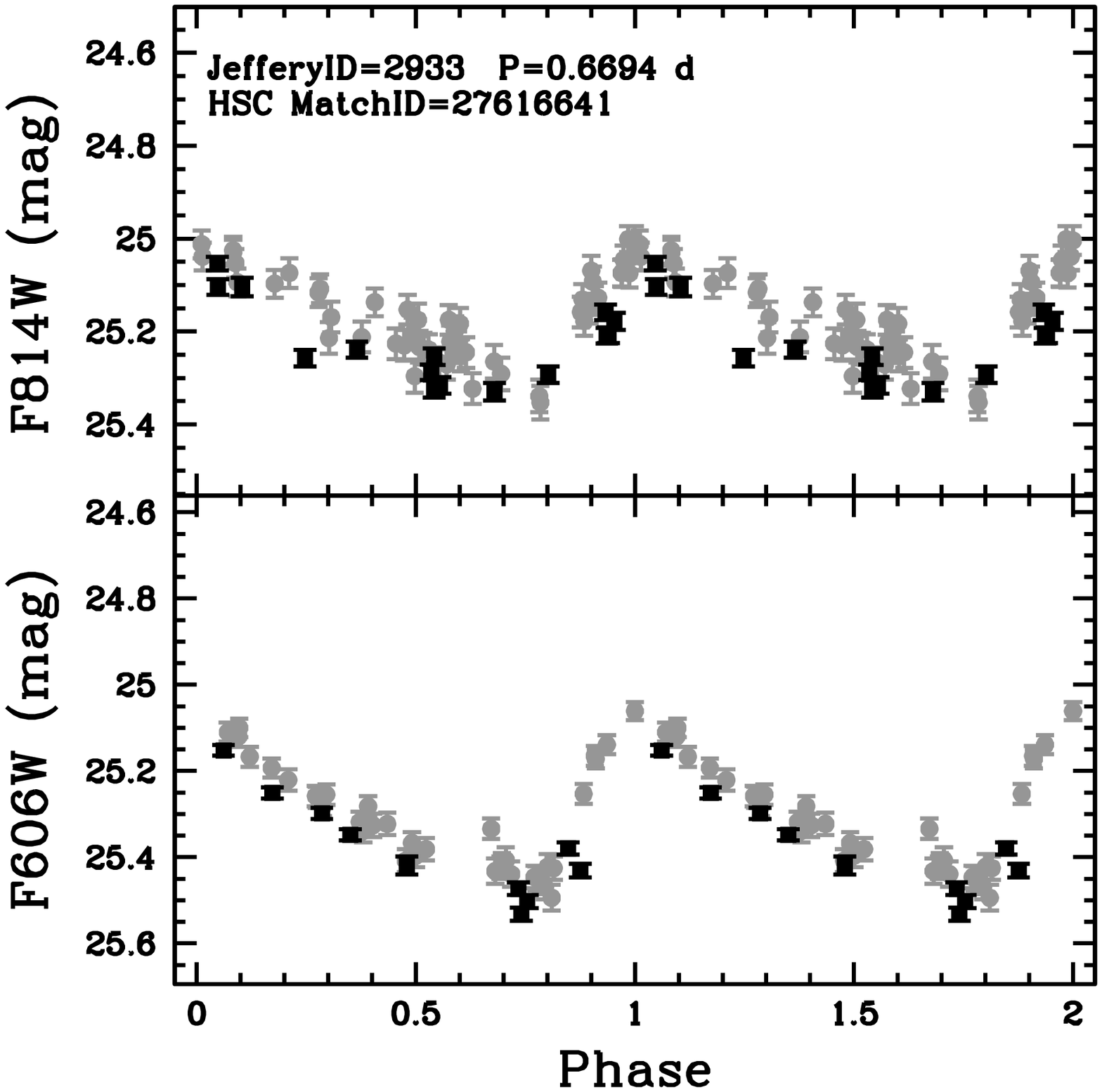} 
\caption{Light curves of two RR~Lyrae stars from HSCv1 (black) and
  literature (grey) from Halo11 (upper panel) and the Stream (lower
  panel). Both are folded with the period provided in the literature.}
\label{fig:LCexample}
\end{centering}
\end{figure}

Figure~\ref{fig:LCexample} shows the LC for two RR~Lyrae variables, one
in the Halo11 and one in the Stream field. Literature LCs in Halo11 are
provided without errors; according to \cite{Brown2004}, they have
typical photometric uncertainty of 0.03\,mag in F606W and 0.04\,mag in
F814W.  The literature data have been converted from STMAG to ABMAG to
match the HSCv1 photometric system, by applying an appropriate aperture
correction using the following equations: ABMAG$_{F606W}={\rm
  STMAG}_{F606W}-0.169+{\rm APC}_{F606W}$, where ${\rm
  APC}_{F606W}=0.248$, and ABMAG$_{F814W}={\rm STMAG}_{F814W}-0.840+{\rm
  APC}_{F814W}$, where ${\rm APC}_{F606W}=0.292$ (\citealt{Brown2009};
see HSCv1 use
case~1\footnote{\url{https://archive.stsci.edu/hst/hsc/help/hsc\_use\_case\_1.html}}).
A small photometric discrepancy of few hundreds of a mag is still
present, likely due to the charge transfer efficiency loss correction
applied by \cite{Brown2009}, but not in HSCv1, as comprehensively
described in \cite{Whitmore2016}.

\begin{table}
 \centering
 \caption{Published variables in the three M\,31 {\it HST} fields.}
 \label{tab:literature}
 \begin{tabular}{l@{~}c@{~~}l@{~}c@{~~}c}
\hline
Field    &  Published  &  Reference  & Listed in & Recovered \\
         &  variables  &	     & HSCv1     & variables \\
\hline
Halo11   & 115  & \protect\cite{Brown2004}     &        88		&      78 (89\%)  \\ 
Disk     &  21  & \protect\cite{Jeffery2011}   &        21		&      21 (100\%) \\  
Stream   &  24  & \protect\cite{Jeffery2011}   &        22		&      21 (95\%)  \\  
\hline
All      & 160  & \protect\cite{Brown2004}     &        131              &  120 (92\%) \\
         &      & \protect\cite{Jeffery2011}   &                         &                 \\
\hline
\end{tabular}
\end{table}

\subsubsection*{Halo11 field}\label{sec:halo11details}

In the Halo11 field, 53 out of 55 published RR~Lyrae variables have a
confirmed counterpart in the HSCv1.  One RR~Lyrae star (V167) has not
been included as it had fewer than 6 data points in the HSCv1 F814W
filter LC. RR~Lyrae variable V124 did not pass the point source
selection, due to blending with a nearby source.

HSCv1 counterparts were confirmed for 27 out of 32 published LPVs (all
32 with amplitudes higher than 0.1 mag in both filters).  Five LPVs
(V14, V18, V19, V21, V25) have no HSCv1 counterpart within 1\,arcsec,
possibly due to their location near the globular cluster SKHB~312
\citep{Holland1997}, where severe blending can be expected.  The
\cite{Brown2004} list of variables also contains two anomalous Cepheid
(aCep\footnote{BLBOO in the GCVS nomenclature
  \citep{2017ARep...61...80S}.})  candidates (V60, V118), one post-AGB
(V84) and two RR~Lyrae star candidates (V29, V6), all successfully
matched within the HSCv1. Due to the shallower magnitude limit of HSCv1,
we could not identify counterparts for the fainter variables published
by \cite{Brown2004}.  Out of 8 eclipsing binary candidates only two have
been matched with HSCv1 (V64 and V173).  Out of 15 dwarf
Cepheids\footnote{Dwarf Cepheids are short-period pulsating variables
  below the Horizontal Branch on the main sequence ($\delta$~Scuti
  stars, DSCT) and SGB (SX~Phoenicis stars, SXPHE).}  only one could be
reliably identified (V48) on the basis of its coordinates and location
on the CMD, although the HSCv1 LC is too noisy to compare reliably with
the published one.

A total of 88 known variables in the Halo11 field (53 confirmed RR~Lyrae
stars and 2 candidates, 27 LPVs, 2 possible anomalous Cepheids, 1 post
AGB, 2 eclipsing binary candidates and 1 possible dwarf Cepheid) have
HSCv1 LCs that pass our selection criteria and are expected to
be recovered by our variability detection technique.  Our procedure has
yielded a total of 156 candidate variables when using both filters, and
128 (167) using only F606W (F814W) data.  Out of the 88 known variables
in the input catalog, 78 (89\%) have been recovered by the variability
selection method when using both filters, 90\% when using only the F606W
filter and 86\% when using only the F814W filter.  The recovery rate is
higher for the F606W data compared to the F814W data, mainly due to the
1.5 times higher amplitude of RR~Lyrae variables in F606W than in F814W.
The variables not recovered are faint (two eclipsing binary candidates,
V64 and V173, and one dwarf Cepheid, V48) or low amplitude variables (a
post-AGB star V84, an LPV 
V158, four low amplitude RR~Lyrae
variables of RRc type, V40, V76, V80 and V137, and one possible RR~Lyrae
variable, V29).  Some of these sources are also affected by problems
such as possible blending, vicinity to the edge of an image, or, to a
diffraction spike of a bright neighboring source.

\subsubsection*{Disk field}
\label{sec:diskdetails}
\cite{Jeffery2011} searched for short period variables discovering 21
RR~Lyrae stars and two Cepheids in the Disk field.  No reliable
counterparts were found in the HSCv1 based sample for two
\cite{Jeffery2011} RRc variables with IDs 15256 (due to a nebulosity)
and 21631 (blending with a brighter source).  We have HSCv1 LCs for 19
known RR~Lyrae variables and 2 Cepheids in the Disk field.

The selection process yielded 192 candidate variables using both
filters, which include all of the 21 known variables, thus yielding a
100\% recovery rate.  For comparison, 266 (202) candidates were found
using only the F606W (F814W) filter, with a recovery rate of 100\%
(90\%) of known variables.

\subsubsection*{Stream field}
\label{sec:streamdetails}
\cite{Jeffery2011} discovered 24 RR~Lyrae stars in the Stream field.
One of the RR~Lyrae stars (ID~409) is missing from the HSCv1 based
sample as it only had 5 visits in F606W filter, probably due to its
location near the edge of the frame. No reliable counterpart was found
for the RR~Lyrae variable ID~8544, probably due to the adverse effect of
a nearby nebulosity on the source extraction procedure employed for
creating the HSCv1. We hence have 22 RR~Lyrae stars in the Stream field
successfully matched with HSCv1.

The variability selection procedure yielded 88 candidate variables using
both filters, out of which 21 are in common with the~\cite{Jeffery2011}
list of 22 known variables in the input catalog (95\% recovery rate).
For comparison, 124 (96) candidates were found using only F606W (F814W)
filter, with a recovery rate of 86\% (91\%).  The only known variable
not recovered is an RRc type RR~Lyrae variable (ID~9975) of low
amplitude ($\simeq$ 0.2\,mag in F606W and $\simeq$0.15 in F814W).

\subsection{New variables}
\label{sec:New}
Candidates  that were not matched 
with known variables were validated using the following procedure:
\begin{enumerate}
\item visual inspection of the images in both filters, in order to
  identify probable sources of error such as: blending; proximity to a
  diffraction spike of another source; cosmic rays; bad column or hot
  pixel; proximity to the frame edge\footnote{Photometry of sources near
  frame edge in the HSCv1 may be affected by incorrect sky background
  estimation.}
    
\item inspection of the LC for the presence of outliers or
  inconsistencies in the appearance of variability in the two filters.

\item lack of obvious similarities with LC of other sources in the
  field, which may indicate systematic errors in photometry.

\end{enumerate}
 The source position on the CMD was used to confirm the variability type
inferred from the LC, but it was not considered as a criterion when
deciding whether a given source is variable or not. 
With the exception of a single source, 22221197, 
all new candidate variables that passed our visual inspection, 
were found to lie in either the RR-Lyrae or the LPV region of the CMD.

Table~\ref{tab:recovery_level2} summarizes the variability search results listing the field
name, the total number of HSCv1 sources passing the initial quality
cuts, the number of candidates obtained with the variable stars
selection procedure, the number of confirmed variables together with the
percentage of recovered known variables, the number of candidates that
were rejected.  
 The process of visual inspection is subjective by its nature. When
presented with apparent low-amplitude variability we had to decide if a
neighboring source is bright and close {\it enough} to corrupt the LC or 
if the LC is {\it sufficiently} different form some other LCs in the field.
For about a half of the rejected candidates listed in Table~\ref{tab:recovery_level2}
we had to make these judgment calls while the other half did not pass 
the saturation-magnitude cuts or had severe imaging problems.

\begin{figure*}
        \includegraphics[width=\textwidth,clip=true,trim=1cm 18cm 1cm 5.5cm]{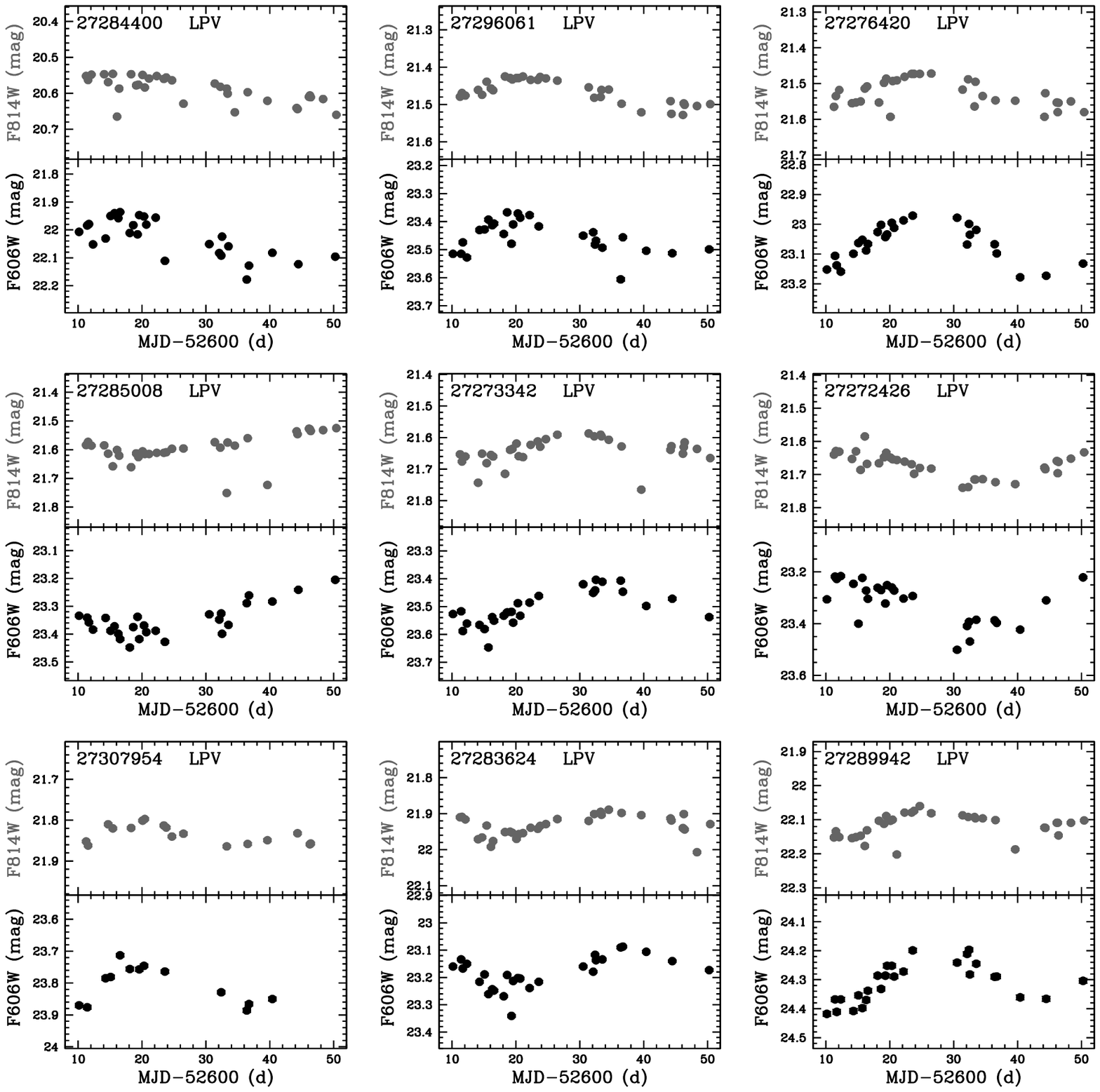}
     \caption{  Sample LCs of the new variables in the Halo11 field. 
 All LC plots are available in Appendix~\ref{sec:appendix}.}
     \label{fig:LCnewSAMPLE}
\end{figure*}

The list of newly identified variables is presented in
Table~\ref{tab:new_variables}.  For each variable the table lists the
HSCv1 ID, Field, equatorial coordinates, F814W average magnitudes,
number of visits in F814W, F606W average magnitudes, number of visits in
F606W, classification and comments.  The LCs of the new
variables are presented in Fig.~\ref{fig:LCnewSAMPLE}.
Fig.~\ref{fig:cmdHalo11}, \ref{fig:cmdDisk} and \ref{fig:cmdStream},
which show the candidates and the confirmed new variables on the CMD
together with all sources.

\begin{table}
 \centering
 \caption{Summary of confirmed and rejected candidate variables.}
 \label{tab:recovery_level2}
 \begin{tabular}{c@{~~}c@{~~}c@{~~}c@{~~}c@{~~}c}
\hline
Field  & HSCv1   & Candidate & Confirmed   	  & Rejected   \\ 
       & sources & variables & candidates         & candidates \\  
\hline
Halo11 &  7109   &    156    & 87 (89\%, 9 new)   &	    69 \\ 
Disk   &  6732   &    192    & 41 (100\%, 20 new) &	   151 \\ 
Stream &  4311   &     88    & 30 (95\%, 9 new)   &	    58 \\ 
\hline
All    &  18152  &    436    & 158 (92\%, 38 new) &        278 \\
\hline
\end{tabular}
\end{table}

\subsubsection*{Halo11 field}
A total of 78 
candidate new variables have been identified. 
Out of these, 13 are rejected because they are brighter than 21.0\,mag in F606W and/or 20.5\,mag in F814W. 
 We visually inspected the images, LC and CMD location of 65 candidates:
30 of these are false detections: 
13 (5 of which are close to globular cluster SKHB~312) 
are affected by blending; 15 are close to the frame edge; 
1 is close to diffraction spikes of a nearby bright star; 1 is surrounded by a halo in some images.  
We were able to confirm 7 new variables, all LPVs.
 The results of visual inspection for the remaining candidates were 
inconclusive, so they were not included in the list of confirmed variables.
 The same procedure was followed for the candidates yielded by the selection 
procedure from the single-filter data resulting in two more confirmed LPVs.

\subsubsection*{Disk field}
A total of 171 candidate new variables were identified using the two
filters.  Out of these, 15 are rejected as being possibly saturated.  We
visually inspected the remaining 156 candidates: 62 are affected by
image problems of several types: 15 are blended with nearby sources, 31
are close to the frame edge, 10 are close to a nebulosity or surrounded
by a halo in some images, 3 are very close diffraction spikes of nearby
sources, 3 are close to hot pixels.  We were able to confirm 20 new
variables: 19 LPVs (one of which at the faint limit, see
Table~\ref{tab:new_variables}) and one with uncertain variability type.
Visual inspection did not result in a firm classification for 74
sources.

\subsubsection*{Stream field}
A total of 67 new candidate variables have been identified using both
filters.  Out of these, 11 have been rejected as possibly saturated.  We
hence visually inspected 56 candidates: 11 are affected by image
problems such as 2 blended, 6 lying on the edge/very close to the edge,
2 very close diffraction spikes of nearby sources, 1 close to hot
pixels.  We were able to confirm 9 new variables, all LPVs (see
Table~\ref{tab:new_variables}).  No other variables could be confirmed
after inspecting candidates selected from the single-filter analysis.

The Stream field observations present a 20 day gap that complicates the
analysis and classification of the LCs.  In this case, we
relied mostly on the shape of the LC in the first part of the
time series.  In some dubious cases (e.g. 27664093, 
see Figure~\ref{fig:LCnewConfirmedStream}), we visually
inspected the LCs for the nearby sources to rule out
photometric problems.

\begin{table*}
 \centering
 \caption{Characteristics of new variable stars ordered by F814W average magnitudes in each field.}
 	\label{tab:new_variables}
 \begin{tabular}{c c c c c c c c c c}
\hline
 MatchID  & Field  & RA (deg)  & Dec (deg) & F814W  & \# Visits & F606W  & \# Visits & Type & Comments\\
  HSCv1   &        & J2000     & J2000     & (mag)  & F814W     & (mag)  & F606W     &      &         \\
\hline
 27284400 & Halo11 & 11.550205 & 40.692210 & 20.586 &  32    & 22.026 &  27    &  LPV &---\\
 27296061 & Halo11 & 11.517188 & 40.681656 & 21.465 &  33    & 23.454 &  27    &  LPV &---\\  
 27276420 & Halo11 & 11.518473 & 40.714756 & 21.528 &  33    & 23.064 &  27    &  LPV &---\\  
 27285008 & Halo11 & 11.553176 & 40.694355 & 21.587 &  31    & 23.353 &  27    &  LPV &---\\  
 27273342 & Halo11 & 11.546703 & 40.740814 & 21.635 &  31    & 23.506 &  27    &  LPV &---\\
 27272426 & Halo11 & 11.552447 & 40.733310 & 21.671 &  33    & 23.316 &  27    &  LPV &---\\  
 27307954 & Halo11 & 11.502041 & 40.674786 & 21.834 &  17    & 23.806 &  13    &  LPV &F606W only\\  
 27283624 & Halo11 & 11.523391 & 40.699898 & 21.932 &  32    & 23.184 &  27    &  LPV &F606W only\\  
 27289942 & Halo11 & 11.504722 & 40.696518 & 22.117 &  33    & 24.310 &  27    &  LPV &---\\  
\hline
22196858  & Disk   & 12.289636 & 42.773460 & 20.882 &  16    & 23.270 &  11    &  LPV &--- \\
22197833  & Disk   & 12.296558 & 42.763630 & 21.333 &  16    & 24.384 &  11    &  LPV &  ---\\    
22211719  & Disk   & 12.291348 & 42.761390 & 21.363 &  16    & 22.572 &  11    &  LPV &--- \\
22206385  & Disk   & 12.283233 & 42.737830 & 21.379 &  16    & 22.310 &  11    &  LPV &  ---\\    
22208821  & Disk   & 12.296735 & 42.758934 & 21.468 &  16    & 23.315 &  11    &  LPV &--- \\
22209236  & Disk   & 12.301403 & 42.769558 & 21.487 &  16    & 23.414 &  11    &  LPV &  ---\\    
22211391  & Disk   & 12.315888 & 42.740982 & 21.618 &  16    & 23.468 &  11    &  LPV &  ---\\    
22210819  & Disk   & 12.296622 & 42.762190 & 21.673 &  16    & 23.573 &  11    &  LPV &  ---\\    
22209784  & Disk   & 12.253314 & 42.765840 & 21.718 &  16    & 24.904 &  11    &  LPV &--- \\
22206501  & Disk   & 12.294667 & 42.731520 & 21.760 &  12    & 25.256 &  11    &  LPV &--- \\
22217682  & Disk   & 12.296862 & 42.739853 & 21.800 &  16    & 23.691 &  11    &  LPV & ---\\
22203253  & Disk   & 12.269516 & 42.742680 & 21.816 &  16    & 23.324 &  11    &  LPV &  ---\\  
22212431  & Disk   & 12.262894 & 42.727220 & 21.826 &  16    & 23.603 &  11    &  LPV & ---\\
22219207  & Disk   & 12.290964 & 42.732960 & 21.832 &  16    & 23.294 &  11    &  LPV & ---\\
22210620  & Disk   & 12.250249 & 42.762253 & 21.845 &  16    & 23.980 &  11    &  LPV & ---\\
22204580  & Disk   & 12.321586 & 42.736850 & 21.890 &  16    & 24.706 &  11    &  LPV &--- \\
22200824  & Disk   & 12.298845 & 42.753952 & 21.922 &  16    & 24.031 &  11    &  LPV & ---\\
22208319  & Disk   & 12.278159 & 42.764660 & 22.163 &  12    & 25.869 &   8    &  LPV &--- \\
22259268  & Disk   & 12.277175 & 42.731316 & 22.977 &  16    & 26.348 &  10    &  LPV & faint limit in F606W\\
22221197  & Disk   & 12.268279 & 42.742805 & 25.637 &  16    & 26.134 &  11    &  other & ---\\
\hline
27663814  & Stream & 11.041388 & 39.788070 & 21.490 &  15    & 23.836 &  10    &  LPV   & ---\\
27661867  & Stream & 11.066272 & 39.778630 & 21.498 &  15    & 24.351 &  10    &  LPV   & --- \\
27654057  & Stream & 11.091002 & 39.781937 & 21.624 &  15    & 23.641 &  10    &  LPV   & ---\\
27659783  & Stream & 11.071672 & 39.786972 & 21.704 &  15    & 24.404 &  9     &  LPV   & ---\\
27664545  & Stream & 11.090615 & 39.765050 & 21.979 &  15    & 23.580 &  10    &  LPV   & --- \\
27665958  & Stream & 11.089661 & 39.768350 & 21.980 &  15    & 24.179 &  10    &  LPV   & ---\\
27664093  & Stream & 11.099528 & 39.772636 & 22.296 &  15    & 24.089 &  10    &  LPV   & ---\\
27630664  & Stream & 11.091218 & 39.797176 & 22.421 &  15    & 23.014 &  10    &  LPV   & ---\\
27619606  & Stream & 11.088532 & 39.802055 & 22.441 &  15    & 24.915 &  10    &  LPV   & ---\\
\hline							   
\end{tabular}
\end{table*}

\section{Discussion}
\label{sec:discussion}
The application of PCA to variability indices that characterize LCs
allowed us to statistically select candidate variable sources using the
admixture coefficients of the first two PCs, $a_1$ and $a_2$. 
The recovery rate of known
variables in the fields studied has been more than $\simeq$90\% in all
cases, while 38 new 
variables were identified. We also
confirmed that various issues that can affect the photometry, such as
saturated pixels, blending, proximity to diffraction spikes, cosmic rays
etc., significantly increase the number of false positives, 
thus making visual inspection of the automatically selected candidates 
imperative. 

 The need for visual inspection of candidate variables is common for all
types of variability search 
\citep[e.g.][]{2010ApJ...712.1259B,2013ApJ...779....7C,2014MNRAS.441.1230A,2014MNRAS.437..132R,2016AJ....151..110K,2016AcA....66..421P}.
The underlying problem is that corrupted measurements cannot be
reliably identified in all cases. 
Individual outlier measurements in a LC can be removed
by sigma-clipping \citep[e.g.][]{Kim2016} or tolerated thanks to
the use of robust variability features \citep{2017A&A...605A.123P}.
However, these measures will not solve the problem if a given object has a
large fraction of its measurements corrupted.
Despite the progress in automatic rejection of imaging artifacts \citep{2002PASP..114..144F,2016A&C....16...67D},
visual inspection of images \citep{2016A&C....16...99M} and LCs remains a valuable quality control tool.
The goal of an automated variability detection algorithm is to minimize the 
percentage of false candidates passed to the visual inspection stage \citep{Fruth2012}.
The PCA-based variability search is doing well in this regard when compared 
to other variability detection techniques applied to the same LC
data (Sec.~\ref{sec:singleidx}).

\subsection{Physical Interpretation of the Principal Components} 
\label{sec:physinterpretationofpc}

\begin{figure*}
\begin{centering}
\includegraphics[width=0.48\textwidth,clip=true,trim=1.3cm 0cm 0.7cm 0cm]{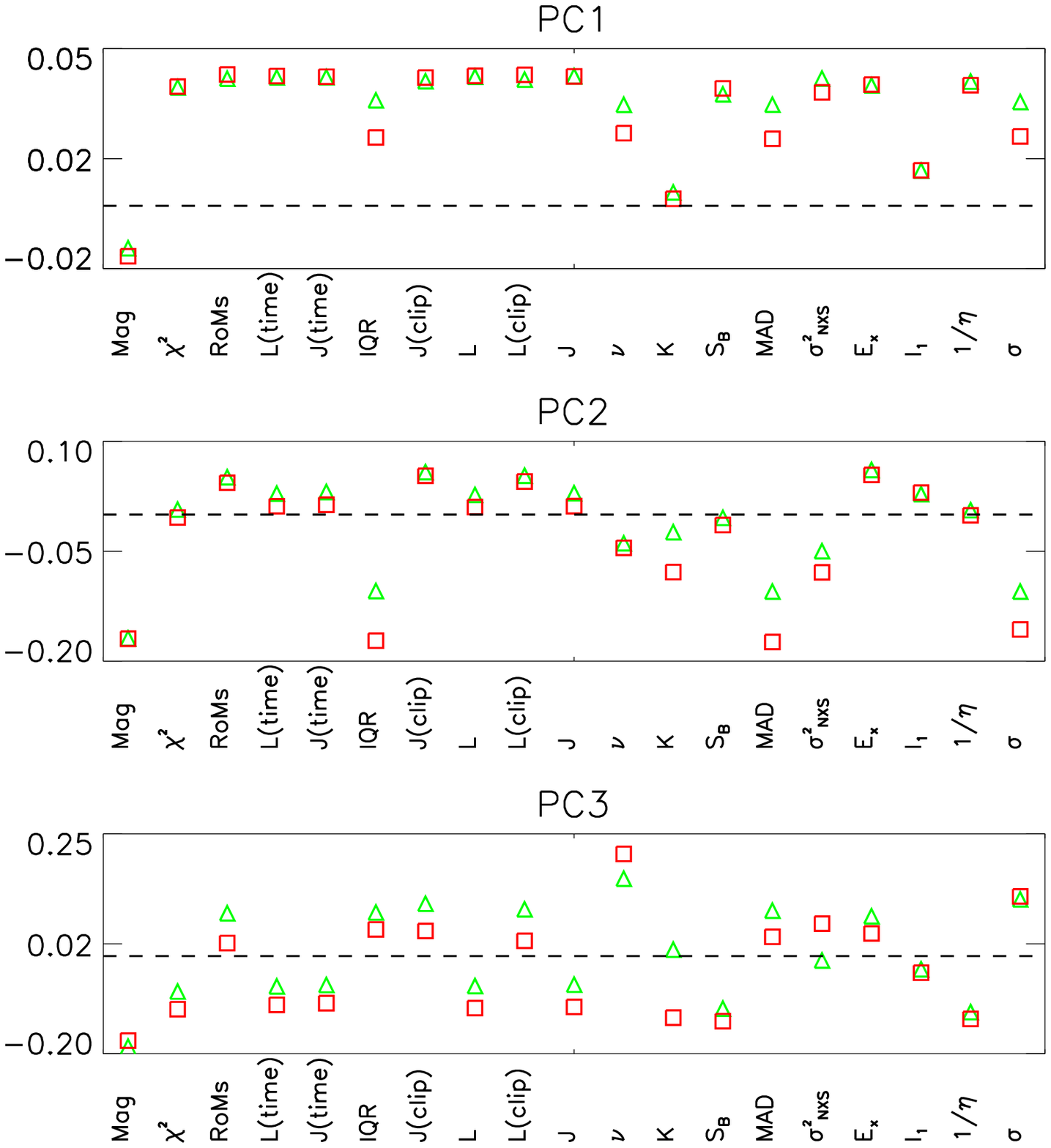}~~~~	
\includegraphics[width=0.48\textwidth,clip=true,trim=1.3cm 0cm 0.7cm 0cm]{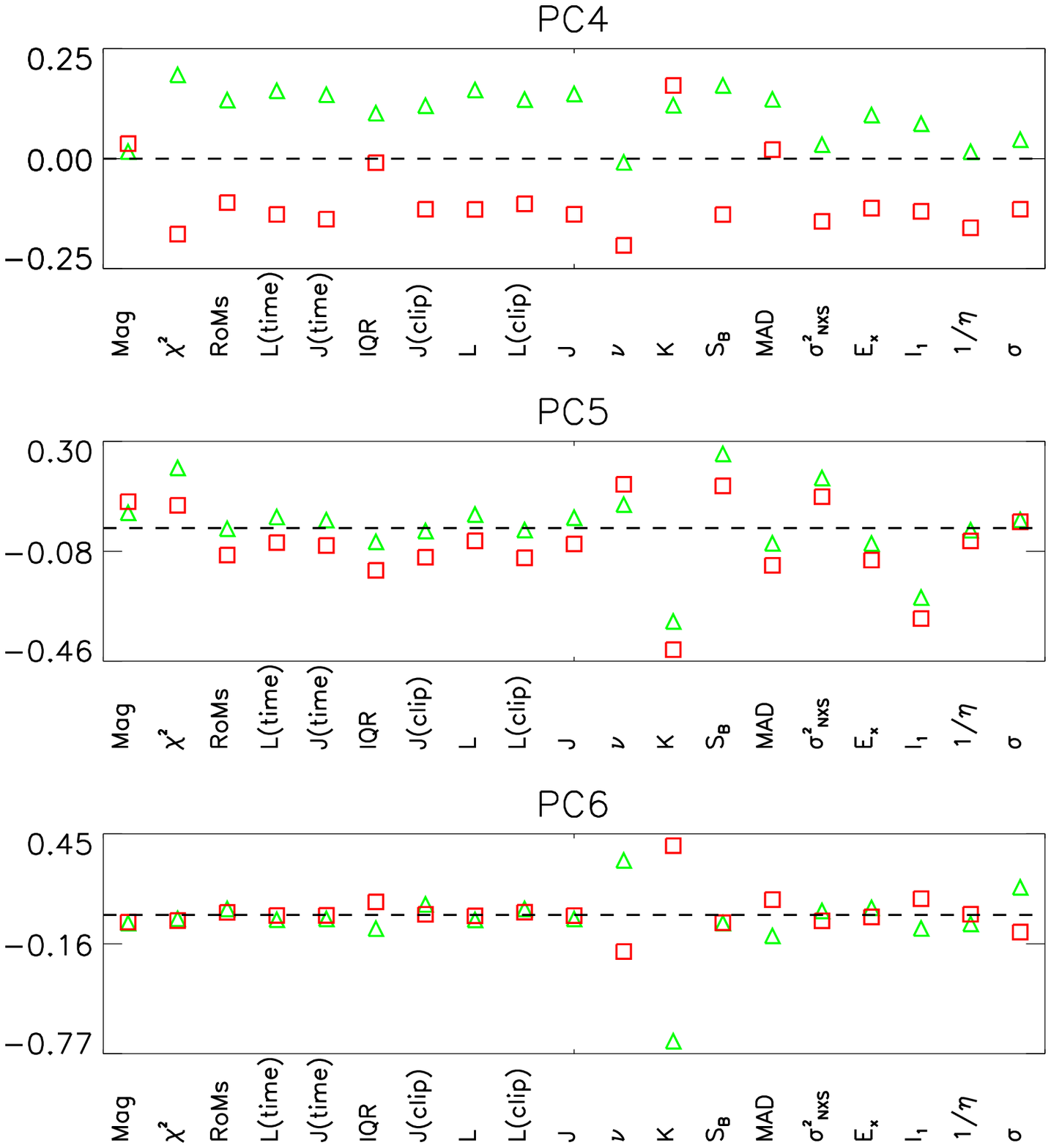}
    \caption{ Coefficients $c_{i,j}$ determining
relative contributions of variability indices to the first six principal
      components for the M\,31 Halo11 data set; see equation~(\ref{eq:pcdefinition}).
      The indices computed using the F606W LCs are shown with green
      triangles and the ones computed using the F814W LCs are shown
      as red squares. The dashed line indicates zero
      contribution of an index to the PC. }
    \label{fig:halo_pcs_both}
\end{centering}
\end{figure*}

 Each PC is a linear combination of all the input variability indices
listed in Table~\ref{tab:indexsummary}:
\begin{equation}
\label{eq:pcdefinition}
\vec{PC_{i}} = 
c_{i,1} \chi_{\rm red}^2 \vec{l_1} + 
c_{i,2} \sigma \vec{l_2}  + 
c_{i,3} {\rm MAD} \vec{l_3} +
c_{i,4} {\rm IQR} \vec{l_4} 
+ \dots
\end{equation}
where $c_{i,1}$, $c_{i,2}$, $c_{i,3}$, $c_{i,4} \dots$ are the coefficients determining
the contribution of each index to the $i$th principal component,
$\vec{PC_{i}}$;
$\vec{l_1}$, $\vec{l_2}$, $\vec{l_3}$, $\vec{l_4} \dots$ are the unit
vectors setting the directions of the $\chi_{\rm red}^2$, $\sigma$, ${\rm MAD}$, ${\rm IQR} \dots$
axes in the variability index space.
As there are different ranges for the numerical values of the variability
indices, they 
are standardized (as described in Sec.~\ref{sec:pcainm31fields})
before entering the equation~(\ref{eq:pcdefinition}).
Figure~\ref{fig:halo_pcs_both} 
presents the contributions, $c_{i,j}$, of  
the variability indices to the first six
principal components 
for the M\,31
Halo11 PCA results.
The dashed line indicates zero contribution of an index to the PC 
 (the values of indices near that line have no effect on the PC value). 
The larger the distance of an index from the zero contribution line, the
more it contributes to this PC. 
 A qualitatively similar relative contribution of the variability indexes to the first few PCs
is found in the Disk and Stream fields (Figures~\ref{fig:disk_pcs_both}, \ref{fig:stream_pcs_both}).

Most variability indices contribute significantly to $\vec{PC_{1}}$. This is
expected, since by construction the indices highlight variability.
$\vec{PC_{1}}$ is
dominated by three scatter-based indices that take into account the
estimated photometric errors ($\chi^2$, RoMS, $\sigma^2_{NXS}$; 
see Table~\ref{tab:indexsummary}) 
and several correlation-based indices
(L, J, S$_B$, E$_X$, $1/\eta$).  The remaining scatter-based indices 
(IQR, MAD, $\sigma$; the first two being robust measures of scatter not
relying on photometric errorbars) contribute to $\vec{PC_{1}}$ as well, 
but they are the main contributors to $\vec{PC_{2}}$ together with the magnitude.
 The admixture coefficient of $\vec{PC_{1}}$, $a_1$, has a large positive value for 
a bright source showing smooth high-amplitude lightcurve. 
The value of $a_2$ is large and negative for a faint 
source with a lightcurve that is not smooth and is showing a large scatter that is
not driven by a few outlier points.
 $\vec{PC_{3}}$--$\vec{PC_{6}}$ show no easily interpretable patterns, while 
together they account for 
15\% of the total data variance.
These components are expected to represent rare information encoded into the variability indices. 
 $K$ and $l_1$ that do not contribute to
$\vec{PC_{1}}$ and $\vec{PC_{2}}$ 
show more significant contribution to
the higher order PCs and in particular $\vec{PC_{5}}$. 
 As discussed in Section~\ref{sec:singleidx}
(Table~\ref{tab:indexcomparison}), the $K$ index (being a robust measure of kurtosis of the distribution of
magnitudes in a lightcurve) cannot identify known variables when used on its 
own \citep[$F_1=0$; c.f.][]{1997ESASP.402..441F}. It does not contribute to 
$\vec{PC_{1}}$ and contributes little to $\vec{PC_{2}}$ (Fig.~\ref{fig:halo_pcs_both}).

It is interesting to note that although in $\vec{PC_{1}}$ and $\vec{PC_{2}}$ the 
variability indices computed in the two filters show similar behavior, 
this is not the case for the higher order PCs. Therefore, the higher order PCs 
highlight differences between LCs in two filters (color changes).

 As the $\vec{PC_{2}}$ is dominated by the indices that are robust measures of
scatter while the correlation-based indices contribute less to the
$\vec{PC_{2}}$,
it becomes possible to separate 
short and long period variables
on the $a_1$--$a_2$ admixture coefficient plots. RR~Lyrae variables 
 (marked with blue circles in Fig.~\ref{fig:halo_admixture})
have
short periods and uncorrelated LCs, for the cadence of the
available observations, thus yielding low absolute values for $a_1$
and
higher values for $a_2$. 
On the contrary, LPVs
 (marked with red squares in Fig.~\ref{fig:halo_admixture})
show smooth LCs with a high degree of correlation between
measurements taken close in time and hence higher values of $a_1$.  As
it can be clearly seen in the 
left panel of
Figure~\ref{fig:halo_admixture} the two types of variables are clearly
separated in the $a_1$--$a_2$ plane.
The two first  PCs are effective not
only in 
separating variable from constant sources, but also in
distinguishing RR~Lyrae variables from LPVs, without the need of
constructing a CMD.
 In general, ``fast'' and ``slow'' (compared to the observing cadence) 
variables occupy separate locations in the $a_1$--$a_2$ plane. There are
only few variables of types other than RR~Lyrae and LPV in the studied
fields to check if further type separation is possible using admixture
coefficients of higher-order PCs.

 Many 
candidate variable sources that were not confirmed 
 by our visual inspection of LCs and images 
occupy a
distinct region on the $a_1$--$a_2$ plane (for the Halo 11 field
$a_1\leq100$ and $a_2\geq12$, cf. Figure ~\ref{fig:halo_admixture}).
 The Disk and Stream fields show a similar location of RR~Lyrae stars, LPVs
and (some) false candidates in the admixture coefficient space (see Appendix).

\subsection{Selection efficiency on the $a_1$--$a_2$ plane}
\label{sec:efficiencytest}
As described in Section~\ref{sec:selection}, the selection process
was performed in two steps. Examination of the resulting candidates from
the first selection step alone, indicated a recovery level of known
variables of $\simeq80$\%, which is significantly lower than the
recovery level achieved when using the fine-tuning process
($\simeq90$\%, for the Halo11 field as an example). On the other hand,
the initial selection yields about half the artifacts selected with
fine-tuning. The artifacts that are not included in the initial
selection are mostly caused by image problems, while the majority of the
artifacts common in both selections are bright sources (probably
affected by saturated pixels).  All new variables were selected as such
in both steps. Therefore, selection of variables using only the first
step described in Section~\ref{sec:selection}, provides lower
completeness (as indicated by the lower recovery rate of known
variables) but higher purity (i.e. lower incidence of
artifacts). Application of the fine-tuning step, provides higher
completeness, but lower purity (thus increasing the number of sources
that need to be expert validated).
 If the number of input lightcurves is very large, one may prefer to use
only the first candidate selection step to reduce the number of candidates
that has to be expected (increase the purity of the candidate list at the
cost of its completeness).

\subsection{Reduced number of source and visits}
\label{sec:sensitivitytest}

The PCA-based variability detection method proposed here is a
statistical method, that requires a substantial sample of sources
(variable and constant) in order to identify candidate variables.  Here
we investigate the applicability of the method, when the size of the
sample is reduced significantly.  We divided the Halo11 sample into
subsamples of different sizes, ranging from 6500 to 500 sources, in
steps of 500. We then applied the PCA variability detection method for
each subsample and calculated the recovery rate of known variables from
\cite{Brown2004} that happen to be included in the specific subsample.
The tests were repeated 100 times for each subsample size, each time
randomly selecting the specified number of sources from the full
sample. The average recovery rate was not affected significantly.  This
test indicates that the method can be applied even to relatively small
samples with as few as 500 sources.

An additional sensitivity test was performed to evaluate the efficiency
of our method when the number of data points in the LC is reduced. The
experiment was conducted using the Halo11 LCs.  Epochs were randomly
omitted from the LCs and the PCA and selection analysis repeated each
time, following the same steps as described in Section~\ref{sec:method}.
The experiment was repeated for 30, 20, 10 and 5 data points in the LC.
 We do not consider LCs with smaller number of points, as the
variability indices (Table~\ref{tab:indexsummary}), while being
mathematically defined, are expected to lose their predictive power.

The decrease of the number of points in the LC expectedly increases the
size of the constant star locus on the $a_1$--$ a_2$ plane, thus
degrading the detectability of variables that lie close to the constant
star locus in the original selection plane of
Figure~\ref{fig:halo_admixture} (upper left panel).  When only 5 points
are retained in the LC, 
additional 10 known variables are not
recovered, reducing the recovery rate to $\simeq$77\%, without
appreciably increasing the false variable detection. This is still an
acceptable result, indicating that the method can be successful even
with a small number of epochs in the LC.

\subsection{Comparison with a single-index search}
\label{sec:singleidx}

\begin{figure} 
\begin{centering}
\includegraphics[width=0.48\textwidth,clip=true,trim=0.0cm 0.0cm 0.0cm 0.0cm]{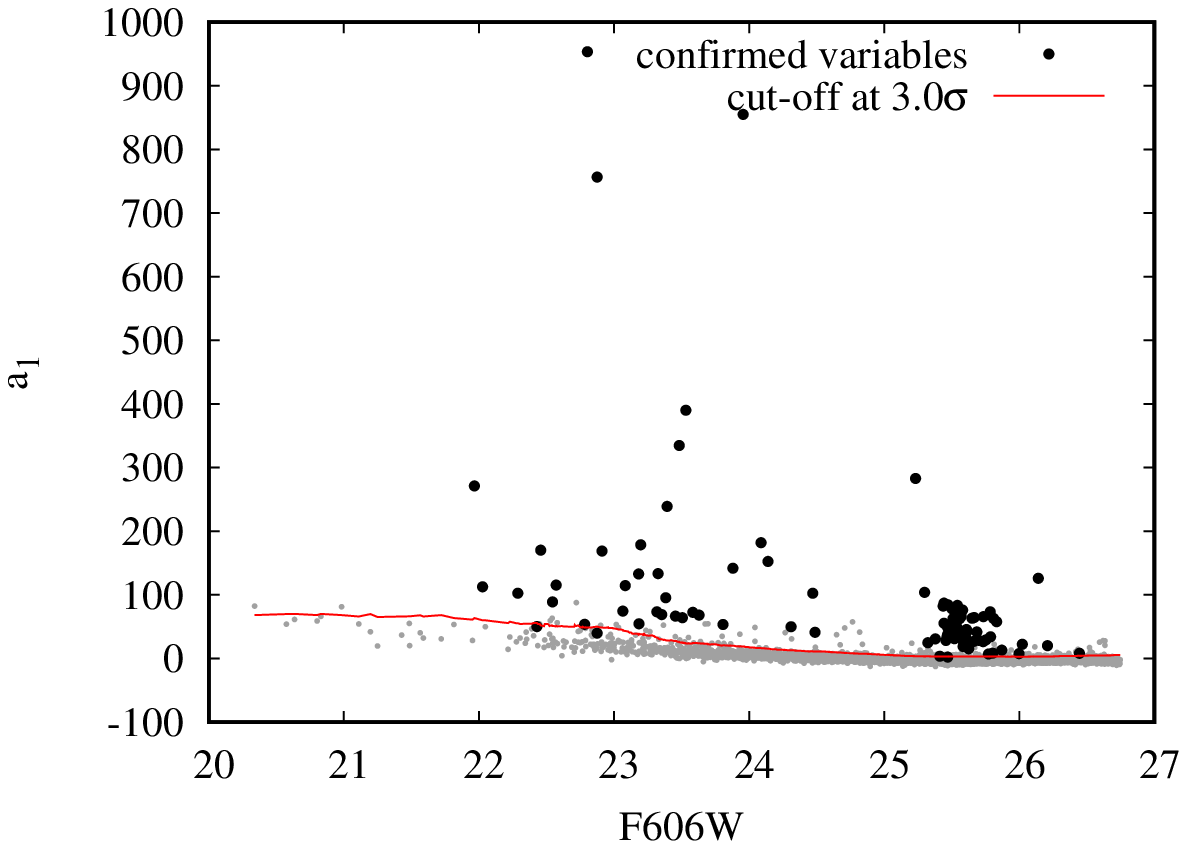}
\includegraphics[width=0.48\textwidth,clip=true,trim=0.0cm 0.0cm 0.0cm 0.0cm]{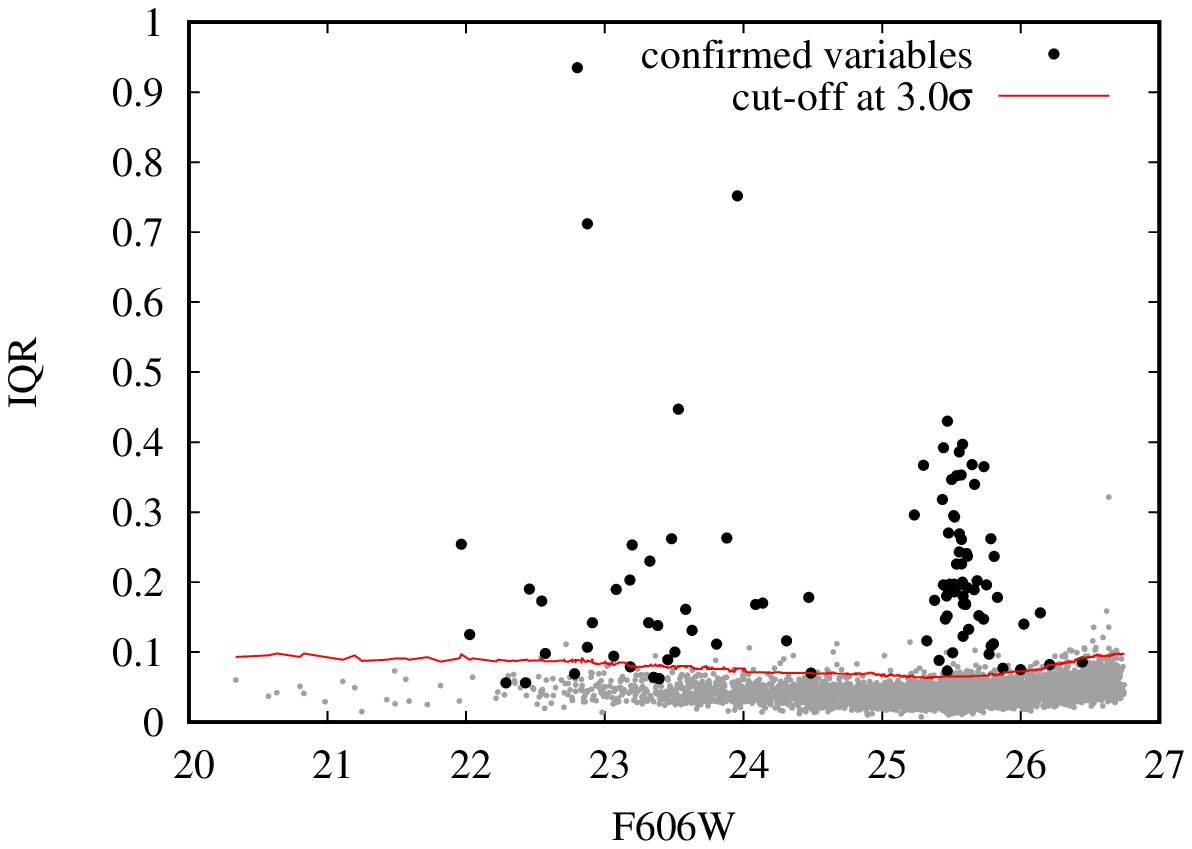}
\includegraphics[width=0.48\textwidth,clip=true,trim=0.0cm 0.0cm 0.0cm 0.0cm]{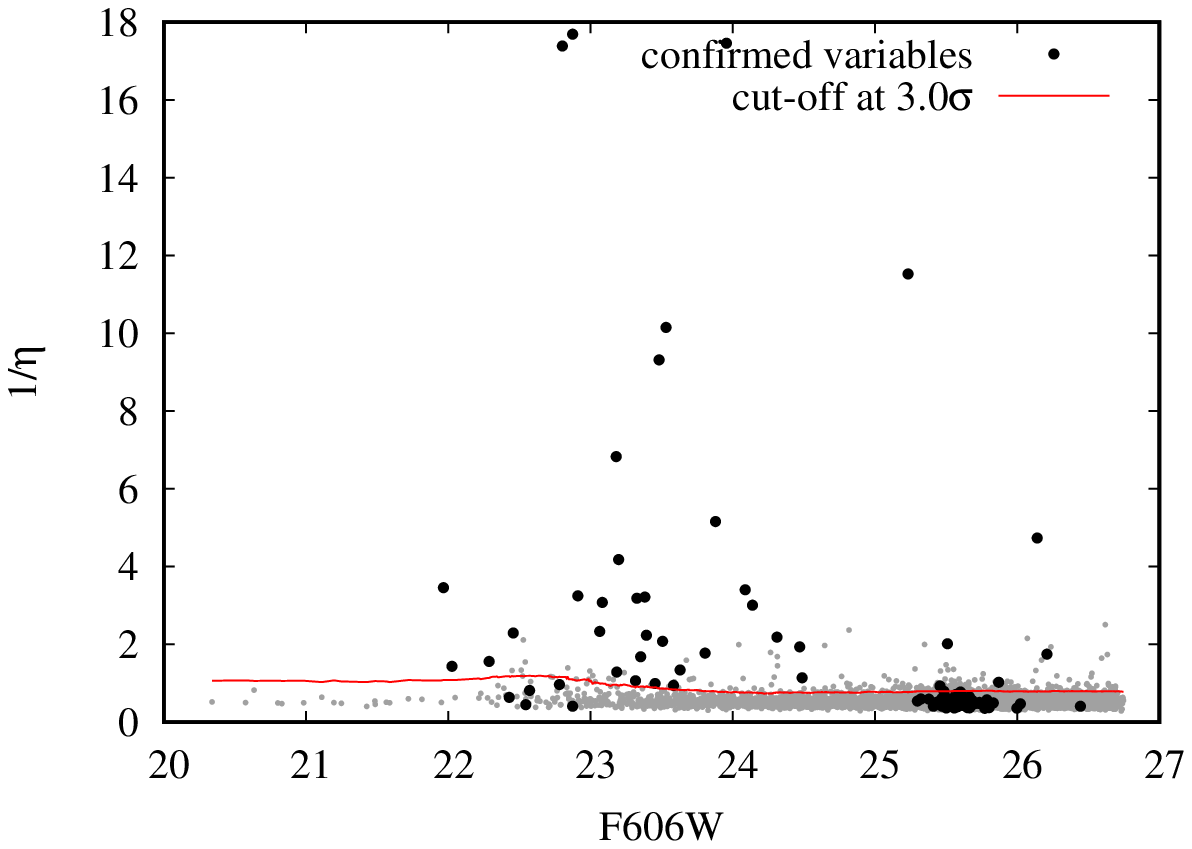} 
\caption{Candidate variable selection in the Halo11 field above magnitude-dependents cuts
(solid line) in $a_1$, ${\rm IQR}$ and $1/\eta$ (Table~\ref{tab:indexsummary}).}
\label{fig:singleindexcuts}
\end{centering}
\end{figure}

 We compare our new variability detection technique relying on
identification of isolated points in the ($a_1$, $a_2$) plane
(presented in Section~\ref{sec:method}) to the conventional methods 
discussed by \cite{Sokolovsky2017}. These methods identify candidate
variables as objects having a value of a single variability index 
(Table~\ref{tab:indexsummary}) above some magnitude-dependent threshold
(Fig.~\ref{fig:singleindexcuts}). The indices are computed independently 
for F606W and F814W LCs.
The authors also suggested to use $a_1$
as a composite variability index. We test this approach
(Fig.~\ref{fig:singleindexcuts}, top panel) combining in $a_1$ 
for each object all the variability indexes computed using its LCs in F606W and F814W bands.

 To evaluate the quality of a variability search in a given set of
LCs we need to know the ground truth: which objects are variable and
which are not. As a working approximation of this we use a list of variables
found by \cite{Brown2004} and \cite{Jeffery2011} cross-matched with the HSC
(Table~\ref{tab:literature}) 
together with the results of our visual inspection of candidate variables
identified with the ($a_1$, $a_2$) technique (Sec.~\ref{sec:New}).
Following \cite{Sokolovsky2017,2017arXiv171007290P} we adopt the
$F_1$-score \citep{vanRijsbergen1974} as the success metric of variability detection.
The $F_1$-score
\begin{equation}
\label{eq:f1}
F_1 = 2 (C \times P) / (C + P)
\end{equation}
is defined through the completeness $C$ (also known as ``recall'') and purity $P$
(``precision'') of the list of
candidate variables \citep{2014MNRAS.439..703G}:
\begin{equation}
\label{eq:completeness}
C = \frac{\mathrm{Number \,\, of \,\, variables \,\, in \,\, the \,\, list}}{\mathrm{Total \,\, number \,\, of \,\, variables \,\, in \,\, the \,\, field}}
\end{equation}
\begin{equation}
\label{eq:purity}
P = \frac{\mathrm{Number \,\, of \,\, variables \,\, in \,\, the \,\, list}}{\mathrm{Total \,\, number \,\, of \,\, candidates \,\, in \,\, the \,\, list}}.
\end{equation}
$F_1=1$ for the perfect selection of variables when one retrieves all
the true variables and no false candidates while $F_1=0$ if no true
variables got into the list of candidates.

\begin{table}
    \caption{Variability detection efficiency ($F_1$-score).}
    \label{tab:indexcomparison}
    \begin{tabular}{r c c c c}
    \hline\hline
  Index                   &    Filter & Halo11 &  Disk &  Stream \\
 \hline
 \multicolumn{5}{c}{PCA-based search} \\
 \hline
($a_1$, $a_2$)  &        & 0.628 &  0.434 &  0.595 \\
$a_1$                &        & 0.564 &  0.373 &  0.530 \\
 \hline
 \multicolumn{5}{c}{Scatter-based indices} \\
 \hline  
$\chi_{\rm red}^2$        &  F606W & 0.419 &  0.154 &  0.260    \\
$\chi_{\rm red}^2$        &  F814W & 0.344 &  0.151 &  0.316    \\
$\sigma$                  &  F606W & 0.532 &  0.230 &  0.412    \\
$\sigma$                  &  F814W & 0.402 &  0.180 &  0.440    \\
${\rm MAD}$               &  F606W & 0.657 &  0.308 &  0.442    \\
${\rm MAD}$               &  F814W & 0.467 &  0.281 &  0.633    \\
${\rm IQR}$               &  F606W & 0.617 &  0.347 &  0.488    \\
${\rm IQR}$               &  F814W & 0.516 &  0.380 &  0.593    \\
${\rm RoMS}$              &  F606W & 0.602 &  0.304 &  0.454    \\
${\rm RoMS}$              &  F814W & 0.528 &  0.322 &  0.547    \\
$\sigma_{\rm NXS}^2$      &  F606W & 0.381 &  0.137 &  0.258    \\
$\sigma_{\rm NXS}^2$      &  F814W & 0.314 &  0.126 &  0.290    \\
$v$                       &  F606W & 0.447 &  0.195 &  0.371    \\
$v$                       &  F814W & 0.350 &  0.146 &  0.323    \\
$K$                       &  F606W & 0.000 &  0.000 &  0.000    \\
$K$                       &  F814W & 0.000 &  0.000 &  0.000    \\
 \hline
 \multicolumn{5}{c}{Correlation-based indices} \\  
 \hline
$J$                       &  F606W & 0.551 &  0.429 &  0.447    \\
$J$                       &  F814W & 0.396 &  0.354 &  0.444    \\
$J({\rm time})$           &  F606W & 0.494 &  0.361 &  0.467    \\
$J({\rm time})$           &  F814W & 0.373 &  0.295 &  0.410    \\
$J({\rm clip})$           &  F606W & 0.671 &  0.381 &  0.542    \\
$J({\rm clip})$           &  F814W & 0.566 &  0.335 &  0.552    \\
$L$                       &  F606W & 0.553 &  0.359 &  0.379    \\
$L$                       &  F814W & 0.404 &  0.336 &  0.354    \\
$L({\rm time})$           &  F606W & 0.481 &  0.372 &  0.467    \\
$L({\rm time})$           &  F814W & 0.391 &  0.298 &  0.378    \\
$L({\rm clip})$           &  F606W & 0.660 &  0.376 &  0.504    \\
$L({\rm clip})$           &  F814W & 0.557 &  0.379 &  0.567    \\
$E_x$                     &  F606W & 0.505 &  0.158 &  0.246    \\
$E_x$                     &  F814W & 0.379 &  0.193 &  0.232    \\
$l_1$                     &  F606W & 0.271 &  0.107 &  0.000    \\
$l_1$                     &  F814W & 0.068 &  0.037 &  0.000    \\
$1/\eta$                  &  F606W & 0.193 &  0.099 &  0.055    \\
$1/\eta$                  &  F814W & 0.176 &  0.091 &  0.056    \\
$S_B$                     &  F606W & 0.257 &  0.099 &  0.146    \\
$S_B$                     &  F814W & 0.232 &  0.104 &  0.121    \\
     \hline
    \end{tabular}
\begin{flushleft}
The $F_1$-score is defined by equation~(\ref{eq:f1}); 
see Table~\ref{tab:indexsummary} for additional information on the
variability indices.
\end{flushleft}
\end{table}

 Table~\ref{tab:indexcomparison} compares the $F_1$-scores reached by
the PCA-based and the conventional single-index techniques in the three
investigated fields. For each individual index we use a $3\sigma$ threshold
to select candidate variables.
Despite a smaller median number of visits (Table~\ref{tab:data}), 
for most individual variability indices the F606W LCs result in higher 
$F_1$-scores than the F814W LCs. This may be attributed to the fact
that pulsating stars tend to have higher variability amplitudes at shorter
wavelengths and are therefore easier to distinguish from non-variable
stars.

 The PCA-based search results in consistently high $F_1$ values compared to
candidate selection based on most individual indices, however the PCA $F_1$ 
is the highest one only in the Disk field. In the Halo11 field, the PCA search is outperformed
by ${\rm MAD}$, $J({\rm clip})$ and $L({\rm clip})$ computed on F606W LCs.
In the Stream field ${\rm MAD}$ computed on the F814W LCs results
in the $F_1$ value higher than the one we obtain with the PCA, while ${\rm IQR}$
(F814W) results in only a slightly lower $F_1$.
The ${\rm MAD}$ computed for F606W LCs results in much lower $F_1$ values
compared to ${\rm MAD}$--F814W in the Stream field, while in the Halo~11
field it is the opposite: ${\rm MAD}$--F606W has the $F_1$-score much higher
than ${\rm MAD}$--F814W.
Our PCA variability search based on two admixture coefficients ($a_1$, $a_2$)
for all three fields results in higher $F_1$ values compared to the PCA-based approach
relying on a magnitude dependent cut in 
$a_1$ considered by \cite{Sokolovsky2017}.

It is hard to predict which index will be
the most efficient variability indicator in a given data set. The efficiency
of an index in identifying variables depends on variability type, observing
cadence, percentage of outlier measurements and the level of correlated
(systematic) noise in the data. The authors suggested ${\rm IQR}$ and $1/\eta$
as the indices that tend to perform well on diverse test data, 
while not necessarily being the best indices for any given data set.
In the Halo11, Disk and Stream data sets considered here, ${\rm IQR}$ is
among the best variability indices, while $1/\eta$ is among the worst ones.
This is easily understood as $1/\eta$ being the measure of correlation
(smoothness) of a lightcurve is unable to detect the numerous RR~Lyrae variables
in these data sets. The observing cadence is long compared to RR~Lyrae
periods, making their LCs appear smooth only when folded with 
the correct period (Fig.~\ref{fig:LCexample}). Only the LPVs which have
smooth LCs plotted as as a function of time can be identified with 
$1/\eta$ (and other correlation-based indices). 

 The PCA-based search results in higher $F_1$ values compared 
to both ${\rm IQR}$ and $1/\eta$ for the three studied fields. 
${\rm MAD}$, another outlier resistant index, may show both higher and
lower $F_1$-scores compared to both ${\rm IQR}$ and the PCA-based search, 
depending of field and filter.
This suggests that, unless it is somehow known a~priory which variability index is the best
one for the studied data set, it is more efficient (in terms of archiving a
higher $F_1$-score) to compute multiple variability indices and combine them 
via the PCA rather than use the ``safe'' indices, ${\rm MAD}$, ${\rm IQR}$ and
$1/\eta$.

 While the PCA constructs a linear combination of indices, 
machine-learning may help to find useful non-linear combinations of indices 
resulting in higher $F_1$-scores 
\citep{2017arXiv171007290P}. 
However, a variability
search based on supervised machine learning requires a representative
training set of LCs pre-classified as variable or non-variable by
some other means and thus cannot be used for a blind variability search in 
a small field.
Training a machine learning classifier on LCs obtained with one 
survey and then applying this classifier to LCs from
another survey (``knowledge transfer''), while being in principle possible,
remains an unsolved problem in practice.
Variability search based on unsupervised machine learning may 
be a promising approach for conducting blind variability surveys 
\citep[e.g.][]{2009MNRAS.400.1897S,2012AJ....143...65S,2016ApJ...820..138M}.

\section{Conclusions}
\label{sec:conclusion}
We investigate a new method of variable object detection in a large set
of light curves, which is based on principal component analysis.  Each
light curve is characterized by its mean magnitude and a set of
variability indices (Table~\ref{tab:indexsummary}) that are used as the
input for the PCA. Candidate variable objects are identified as outliers
in the plane of admixture coefficients $a_1$--$a_2$ corresponding to the
two most significant principal components. The proposed method is
suitable for (and most efficient in) large sets of LCs.  It
requires no a~priori information about the type of the searched variable
objects. Instead it relies on assumptions that variable objects are rare
and the indices listed in Table~\ref{tab:indexsummary} are able to
capture variability information -- the assumptions that hold true for
most ground-based and many space-based observations in optical and
near-infrared bands. This methodology can indeed successfully identify variable stars not only
in the HSC database, but also in ground based data
with different sample sizes and epochs as discussed in Section~\ref{sec:sensitivitytest}.
There is no need to preselect the ``best''
variability indices for different samples, since PCA outputs their
optimal linear combinations.  Human intervention is still necessary to
validate the resulting candidate variables (see discussion in~\ref{sec:efficiencytest}). 
The present algorithm is performed in the framework of the HCV, 
which will be available in 2018.

The method is verified using 18152 LCs of stars in 3 fields in
M\,31 extracted from the {\it Hubble} Source Catalogue. We recovered
about 90\% of the known variables reported by \citet{Brown2004} and
\citet{Jeffery2011}.  We found 38 new variable stars, among which 37
LPVs and one object of an uncertain variability type
(Table~\ref{tab:new_variables}).  This demonstrates that the {\it
  Hubble} Source Catalogue, despite its shallower depth and reduced time
resolution compared to what can be obtained with dedicated manual
analysis of the archival {\it HST} images, includes many previously
unknown variable objects in fields previously studied for variability.

\section*{Acknowledgements}
We acknowledge financial support by the European Space Agency (ESA)
under the ``{\it Hubble} Catalog of Variables'' program, contract
No.~4000112940.  This work uses the HSC, based on observations made with
the NASA/ESA {\it Hubble Space Telescope} and obtained from the {\it
  Hubble} Legacy Archive, which is a collaboration between the Space
Telescope Science Institute (STScI/NASA), the Space Telescope European
Coordinating Facility (ST-ECF/ESAC/ESA) and the Canadian Astronomy Data
Centre (CADC/NRC/CSA).  This research has made use of NASA's
Astrophysics Data System. We acknowledge support
of the HCV by Danny Lennon and Brad Whitmore (B.W.), useful discussions 
by B.W., Ioannis Bellas-Velidis, Ektoras Pouliasis, Zoi Spetsieri and 
critical comments on the manuscript by Vassilis Charmandaris.

%%%%%%%%%%%%%%%%%%%%%%%%%%%%%%%%%%%%%%%%%%%%%%%%%%



%\bibliographystyle{mnras}

\begin{thebibliography}{199}

\bibitem[Abbas et al.(2014)]{2014MNRAS.441.1230A} Abbas M.~A., Grebel E.~K., Martin N.~F., Burgett W.~S., Flewelling H., Wainscoat R.~J., 2014, MNRAS, 441, 1230 
\bibitem[Ahn et al.(2014)]{2014ApJS..211...17A} Ahn C.~P., et al., 2014, ApJS, 211, 17 
\bibitem[Alonso et al.(2007)]{Alonso2007} Alonso, R., Brown, T.~M., Charbonneau, D., et al.\ 2007, Transiting Extrasolar Planets Workshop, 366, 13 
\bibitem[Angeloni et al.(2014)]{2014A&A...567A.100A} Angeloni R., et al., 2014, A\&A, 567, A100 
\bibitem[Auvergne et al.(2009)]{Auvergne2009} Auvergne, M., Bodin, P., Boisnard, L., et al.\ 2009, \aap, 506, 411
\bibitem[Bailer-Jones et al. (1998)]{Bailer1998} Bailer-Jones, C.~A.~L., Irwin, M., \& von Hippel, T.\ 1998, \mnras, 298, 361
\bibitem[Bakos et al.(2004)]{Bakos2004} Bakos, G., Noyes, R.~W., Kov{\'a}cs, G., et al.\ 2004, \pasp, 116, 266 
\bibitem[Benedict et al.(2017)]{2017PASP..129a2001B} Benedict G.~F., McArthur B.~E., Nelan E.~P., Harrison T.~E., 2017, PASP, 129, 012001 
\bibitem[Bernard et al.(2010)]{2010ApJ...712.1259B} Bernard E.~J., et al., 2010, ApJ, 712, 1259 
\bibitem[Bernard et al.(2013)]{2013MNRAS.432.3047B} Bernard E.~J., et al., 2013, MNRAS, 432, 3047 
\bibitem[Bertin \& Arnouts(1996)]{BertinAnouts1996} Bertin, E., \& Arnouts, S.\ 1996, \aaps, 117, 393 
\bibitem[Bonanos \& Stanek(2003)]{Bonanos2003} Bonanos, A.~Z., \& Stanek, K.~Z.\ 2003, \apjl, 591, L111
\bibitem[Borucki et al.(2010)]{Borucki2010} Borucki, W.~J., Koch, D., Basri, G., et al.\ 2010, Science, 327, 977 
\bibitem[Brown et al.(1989)]{Brown1989} Brown, L.~M.~J., Robson, E.~I., Gear, W.~K., \& Smith, M.~G.\ 1989, \apj, 340, 150
\bibitem[Brown et al.(2003)]{Brown2003} Brown, T.~M., Ferguson, H.~C., Smith, E., et al.\ 2003, \apjl, 592, L17
\bibitem[Brown et al.(2004)]{Brown2004} Brown, T.~M., Ferguson, H.~C., Smith, E., et al.\ 2004, \aj, 127, 2738
\bibitem[Brown et al.(2006)]{Brown2006} Brown, T.~M., Smith, E., Ferguson, H.~C., et al.\ 2006, \apj, 652, 323
\bibitem[Brown et al.(2009)]{Brown2009} Brown, T.~M., Smith, E., Ferguson, H.~C., et al.\ 2009, \apjs, 184, 152
\bibitem[Budav{\'a}ri \& Lubow(2012)]{BudavariLubow2012} Budav{\'a}ri, T., \& Lubow, S.~H.\ 2012, \apj, 761, 188 
\bibitem[Burdanov et al.(2016)]{2016MNRAS.461.3854B} Burdanov A.~Y., et al., 2016, MNRAS, 461, 3854 
\bibitem[Butters et al.(2010)]{Butters2010} Butters, O.~W., West, R.~G., Anderson, D.~R., et al.\ 2010, \aap, 520, L10
\bibitem[Chambers et al.(2016)]{2016arXiv161205560C} Chambers K.~C., et al., 2016, arXiv:1612.05560 
\bibitem[Chazelas et al.(2012)]{2012SPIE.8444E..0EC} Chazelas B., et al., 2012, SPIE, 8444, 84440E 
\bibitem[Christ, Kempa-Liehr, \& Feindt(2016)]{2016arXiv161007717C} Christ M., Kempa-Liehr A.~W., Feindt M., 2016, arXiv:1610.07717 
\bibitem[Cioni et al.(2011)]{Cioni2011} Cioni M.-R.~L., Clementini G., Girardi L., et al.\ 2011, \aap, 527, A116 
\bibitem[Clementini et al.(2009)]{Clementini2009} Clementini, G., Contreras, R., Federici, L., et al.\ 2009, \apjl, 704, L103
\bibitem[Clementini et al.(2016)]{Clementini2016} Clementini, G., Ripepi, V., Leccia, S., et al.\ 2016, \aap, 595, A133 
\bibitem[Cusano et al.(2013)]{2013ApJ...779....7C} Cusano F., et al., 2013, ApJ, 779, 7 
\bibitem[Debosscher et al.(2007)]{Debosscher2007} Debosscher, J., Sarro, L.~M., Aerts, C., et al.\ 2007, \aap, 475, 1159
\bibitem[Debosscher et al.(2009)]{Debosscher2009} Debosscher, J., Sarro, L.~M., L{\'o}pez, M., et al.\ 2009, \aap, 506, 519
\bibitem[de Diego(2010)]{deDiego2010} de Diego, J.~A.\ 2010, \aj, 139, 1269
\bibitem[Desai et al.(2016)]{2016A&C....16...67D} Desai S., Mohr J.~J., Bertin E., K{\"u}mmel M., Wetzstein M., 2016, A\&C, 16, 67 
\bibitem[de Souza et al.(2014)]{deSouza2014} de Souza, R.~S., Maio, U., Biffi, V., \& Ciardi, B.\ 2014, \mnras, 440, 240
\bibitem[Di Criscienzo et al.(2011)]{DiCriscienzo2011} Di Criscienzo, M., Greco, C., Ripepi, V., et al.\ 2011, \aj, 141, 81
\bibitem[Drake et al.(2009)]{Drake2009} Drake, A.~J., Djorgovski, S.~G., Mahabal, A., et al.\ 2009, \apj, 696, 870 
\bibitem[Drake et al.(2014)]{2014ApJS..213....9D} Drake A.~J., et al., 2014, ApJS, 213, 9 
\bibitem[Dutta et al.(2018)]{2018arXiv180202303D} Dutta S., Mondal S., Joshi S., Jose J., Das R., Ghosh S., 2018, arXiv:1802.02303 
\bibitem[Eyer(2006)]{Eyer2006} Eyer, L.\ 2006, Astrophysics of Variable Stars, 349, 15 
\bibitem[Ferreira Lopes \& Cross(2016a)]{FerreiraCross2016} Ferreira Lopes, C.~E., \& Cross, N.~J.~G.\ 2016, \aap, 586, A36
\bibitem[Ferreira Lopes \& Cross(2017)]{FerreiraCross2017} Ferreira Lopes, C.~E., \& Cross, N.~J.~G.\ 2017, \aap, 604, A121 
\bibitem[Figuera Jaimes et al.(2013)]{Figuera2013} Figuera Jaimes, R., Arellano Ferro, A., Bramich, D.~M., Giridhar, S., \& Kuppuswamy, K.\ 2013, \aap, 556, A20
\bibitem[Fiorentino et al.(2013)]{Fiorentino2013} Fiorentino, G., Musella, I., \& Marconi, M.\ 2013, \mnras, 434, 2866
\bibitem[Fiorentino et al.(2010)]{Fiorentino2010} Fiorentino, G., Contreras Ramos, R., Clementini, G., et al.\ 2010, \apj, 711, 808
\bibitem[Freedman et al.(2001)]{Freedman2001} Freedman, W.~L., Madore, B.~F., Gibson, B.~K., et al.\ 2001, \apj, 553, 47
\bibitem[Friedrich, Koenig, \& Wicenec(1997)]{1997ESASP.402..441F} Friedrich S., Koenig M., Wicenec A., 1997, ESASP, 402, 441 
\bibitem[Fruchter \& Hook(2002)]{2002PASP..114..144F} Fruchter A.~S., Hook R.~N., 2002, PASP, 114, 144 
\bibitem[Fruth et al.(2012)]{Fruth2012} Fruth, T., Kabath, P., Cabrera, J., et al.\ 2012, \aj, 143, 140
\bibitem[Gaia Collaboration et al.(2016)]{GaiaCollaboration2016} Gaia Collaboration, Prusti, T., de Bruijne, J.~H.~J., et al.\ 2016, \aap, 595, A1 
\bibitem[Gavras et al.(2017)]{Gavras2017} Gavras, P., Bonanos, A.~Z., Bellas-Velidis, I., et al.\ 2017, IAU Symposium, 325, 369
\bibitem[Graham et al.(2014)]{2014MNRAS.439..703G} Graham M.~J., Djorgovski S.~G., Drake A.~J., Mahabal A.~A., Chang M., Stern D., Donalek C., Glikman E., 2014, MNRAS, 439, 703 
\bibitem[Griest et al.(1991)]{Griest1991} Griest, K., Alcock, C., Axelrod, T.~S., et al.\ 1991, \apjl, 372, L79
\bibitem[Graham et al.(2013)]{Graham2013} Graham, M.~J., Drake, A.~J., Djorgovski, S.~G., et al.\ 2013, \mnras, 434, 3423
\bibitem[Hoffmann et al.(2016)]{2016ApJ...830...10H} Hoffmann S.~L., et al., 2016, ApJ, 830, 10 
\bibitem[Holland et al.(1997)]{Holland1997} Holland, S., Fahlman, G.~G., \& Richer, H.~B.\ 1997, \aj, 114, 1488
\bibitem[Horne \& Baliunas(1986)]{HorneBaliunas1986} Horne, J.~H., \& Baliunas, S.~L.\ 1986, \apj, 302, 757
\bibitem[Ibata et al.(2001)]{Ibata2001} Ibata, R., Irwin, M., Lewis, G., Ferguson, A.~M.~N., \& Tanvir, N.\ 2001, \nat, 412, 49 
\bibitem[Ishida \& de Souza(2013)]{Ishida2013} Ishida, E.~E.~O., \& de Souza, R.~S.\ 2013, \mnras, 430, 509
\bibitem[Ivezic et al.(2008)]{2008arXiv0805.2366I} Ivezic Z., et al., 2008, arXiv:0805.2366 
\bibitem[Jeffery et al.(2011)]{Jeffery2011} Jeffery, E.~J., Smith, E., Brown, T.~M., et al.\ 2011, \aj, 141, 171
\bibitem[Karampelas et al. (2012)]{Karampelas2012} Karampelas, A., Kontizas, M., Rocca-Volmerange, B., et al.\ 2012, \aap, 538, A38
\bibitem[Kim et al.(2011)]{Kim2011} Kim, D.-W., Protopapas, P., Alcock, C., Byun, Y.-I., \& Khardon, R.\ 2011, Astronomical Data Analysis Software and Systems XX, 442, 447
\bibitem[Kim et al.(2014)]{Kim2014} Kim, D.-W., Protopapas, P., Bailer-Jones, C.~A.~L., et al.\ 2014, \aap, 566, A43
\bibitem[Kim \& Bailer-Jones(2016)]{Kim2016} Kim, D.-W., \& Bailer-Jones, C.~A.~L.\ 2016, \aap, 587, A18 
\bibitem[Kirihara et al.(2017)]{Kirihara2017} Kirihara, T., Miki, Y., Mori, M., Kawaguchi, T., \& Rich, R.~M.\ 2017, \mnras, 464, 3509 
\bibitem[Klagyivik et al.(2016)]{2016AJ....151..110K} Klagyivik P., et al., 2016, AJ, 151, 110 
\bibitem[Koch et al. (2010)]{Koch2010} Koch, D.G., Borucki, W.J., Basri, G. et al., 2010, \apj, 713, 79
\bibitem[Kolesnikova et al.(2008)]{Kolesnikova2008} Kolesnikova, D.~M., Sat, L.~A., Sokolovsky, K.~V., Antipin, S.~V., \& Samus, N.~N.\ 2008, \actaa, 58, 279
\bibitem[K{\"u}gler, Gianniotis, \& Polstere(2015)]{2015MNRAS.451.3385K} K{\"u}gler S.~D., Gianniotis N., Polsterer K.~L., 2015, MNRAS, 451, 3385 
\bibitem[Laher et al.(2017)]{2017arXiv170801584L} Laher R.~R., et al., 2017, arXiv:1708.01584 
\bibitem[Law et al.(2009)]{Law2009} Law, N.~M., Kulkarni, S.~R., Dekany, R.~G., et al.\ 2009, \pasp, 121, 1395
\bibitem[Lasker et al.(2008)]{Lasker2008} Lasker, B.~M., Lattanzi, M.~G., McLean, B.~J., et al.\ 2008, \aj, 136, 735 
\bibitem[Layden et al.(1999)]{1999AJ....117.1313L} Layden A.~C., Ritter L.~A., Welch D.~L., Webb T.~M.~A., 1999, AJ, 117, 1313 
\bibitem[Mackenzie, Pichara, \& Protopapas(2016)]{2016ApJ...820..138M} Mackenzie C., Pichara K., Protopapas P., 2016, ApJ, 820, 138 
\bibitem[Medina et al.(2018)]{2018arXiv180201581M} Medina G., et al., 2018, arXiv, arXiv:1802.01581 
\bibitem[McCommas et al.(2009)]{2009AJ....137.4707M} McCommas L.~P., Yoachim P., Williams B.~F., Dalcanton J.~J., Davis M.~R., Dolphin A.~E., 2009, AJ, 137, 4707 
\bibitem[Melchior et al.(2016)]{2016A&C....16...99M} Melchior P., et al., 2016, A\&C, 16, 99 
\bibitem[Minniti et al.(2010)]{Minniti2010} Minniti, D., Lucas, P.~W., Emerson, J.~P., et al.\ 2010, \na, 15, 433
\bibitem[Mowlavi(2014)]{2014A&A...568A..78M} Mowlavi N., 2014, A\&A, 568, A78 
\bibitem[Nandra et al.(1997)]{Nandra1997} Nandra, K., George, I.~M., Mushotzky, R.~F., Turner, T.~J., \& Yaqoob, T.\ 1997, \apj, 476, 70
\bibitem[Nun et al.(2015)]{Nun2015} Nun, I., Protopapas, P., Sim, B., et al.\ 2015, arXiv:1506.00010
\bibitem[Oelkers et al.(2018)]{2018AJ....155...39O} Oelkers R.~J., et al., 2018, AJ, 155, 39 
\bibitem[Paegert et al.(2014)]{Paegert2014} Paegert, M., Stassun, K.~G., \& Burger, D.~M.\ 2014, \aj, 148, 31
\bibitem[Parks et al.(2014)]{Parks2014} Parks, J.~R., Plavchan, P., White, R.~J., \& Gee, A.~H.\ 2014, \apjs, 211, 3
\bibitem[Pashchenko, Sokolovsky, \& Gavras.(2017)]{2017arXiv171007290P} Pashchenko I.~N., Sokolovsky K.~V., Gavras P., 2017, arXiv:1710.07290 
\bibitem[Pawlak et al.(2016)]{2016AcA....66..421P} Pawlak M., et al., 2016, AcA, 66, 421 
\bibitem[Pearson(1901)]{Pearson1901} Pearson K., 1901, Philosophical Magazine Series 6, 2, 11, 559
\bibitem[Pepper et al.(2007)]{Pepper2007} Pepper, J., Pogge, R.~W., DePoy, D.~L., et al.\ 2007, \pasp, 119, 923 
\bibitem[P{\'e}rez-Ortiz et al.(2017)]{2017A&A...605A.123P} P{\'e}rez-Ortiz M.~F., Garc{\'{\i}}a-Varela A., Quiroz A.~J., Sabogal B.~E., Hern{\'a}ndez J., 2017, A\&A, 605, A123 
\bibitem[Perryman et al.(2001)]{Perryman2001} Perryman, M.~A.~C., de Boer, K.~S., Gilmore, G., et al.\ 2001, \aap, 369, 339
\bibitem[Pojmanski(2002)]{Pojmanski2002} Pojmanski, G.\ 2002, \actaa, 52, 397 
\bibitem[Re Fiorentin et al. (2007)]{Fiorentin2007} Re Fiorentin, P., Bailer-Jones, C.~A.~L., Lee, Y.~S., et al.\ 2007, \aap, 467, 1373
\bibitem[Ramsay et al.(2014)]{2014MNRAS.437..132R} Ramsay G., et al., 2014, MNRAS, 437, 132 
\bibitem[Richards et al.(2011)]{Richards2011} Richards, J.~W., Starr, D.~L., Butler, N.~R., et al.\ 2011, \apj, 733, 10 
\bibitem[Ricker et al.(2014)]{2014SPIE.9143E..20R} Ricker G.~R., et al., 2014, SPIE, 9143, 914320 
\bibitem[Rose \& Hintz(2007)]{Rose&Hintz2007} Rose, M.~B., \& Hintz, E.~G.\ 2007, \aj, 134, 2067
\bibitem[Samus et al.(2017)]{2017ARep...61...80S} Samus N.~N., Kazarovets E.~V., Durlevich O.~V., Kireeva N.~N., Pastukhova E.~N., 2017, ARep, 61, 80
\bibitem[Sesar et al.(2017)]{2017AJ....153..204S} Sesar B., et al., 2017, AJ, 153, 204 
\bibitem[Shappee et al. (2014)]{Shappee2014} Shappee, B.J., Prieto, J.L., Grupe, D., et al., 2014, \apj, 788, 48
\bibitem[Shin et al.(2009)]{2009MNRAS.400.1897S} Shin, M.-S., Sekora, M., \& Byun, Y.-I.\ 2009, \mnras, 400, 1897
\bibitem[Shin et al.(2012)]{2012AJ....143...65S} Shin M.-S., Yi H., Kim D.-W., Chang S.-W., Byun Y.-I., 2012, AJ, 143, 65 
\bibitem[Skrutskie et al.(2006)]{2006AJ....131.1163S} Skrutskie M.~F., et al., 2006, AJ, 131, 1163 
\bibitem[Sokolovsky et al.(2017a)]{Sokolovsky2017} Sokolovsky, K.~V., Gavras, P., Karampelas, A., et al.\ 2017, \mnras, 464, 274
\bibitem[Sokolovsky et al.(2017b)]{Sokolovsky2017b} Sokolovsky, K., Bonanos, A., Gavras, P., et al.\ 2017, arXiv:1703.02038
\bibitem[Sokolovsky \& Lebedev(2018)]{2018A&C....22...28S} Sokolovsky K.~V., Lebedev A.~A., 2018, A\&C, 22, 28 
\bibitem[Stetson(1996)]{Stetson1996} Stetson, P.~B.\ 1996, \pasp, 108, 851
\bibitem[Steiner et al.(2009)]{Steiner2009} Steiner, J.~E., Menezes, R.~B., Ricci, T.~V., \& Oliveira, A.~S.\ 2009, \mnras, 395, 64
\bibitem[S{\"u}veges et al.(2012)]{Suveges2012} S{\"u}veges, M., Sesar, B., V{\'a}radi, M., et al.\ 2012, \mnras, 424, 2528 
\bibitem[Tisserand et al.(2007)]{Tisserand2007} Tisserand P., et al.\ 2007, \aap, 469, 387 
\bibitem[Udalski et al.(2008)]{Udalski2008} Udalski, A., Szymanski, M.~K., Soszynski, I., \& Poleski, R.\ 2008, \actaa, 58, 69
\bibitem[van~Rijsbergen(1974)]{vanRijsbergen1974}van~Rijsbergen C.~J., 1974, Journal of Documentation, 30, 12, 365
\bibitem[Welch \& Stetson(1993)]{1993AJ....105.1813W} Welch D.~L., Stetson P.~B., 1993, AJ, 105, 1813 
\bibitem[Wheatley et al.(2017)]{2017arXiv171011100W} Wheatley P.~J., et al., 2017, arXiv:1710.11100 
\bibitem[Whitmore et al.(2016)]{Whitmore2016} Whitmore B.~C., et al., 2016, AJ, 151, 134 
\bibitem[Wo{\'z}niak et al.(2004)]{Wozniak2004} Wo{\'z}niak, P.~R., Vestrand, W.~T., Akerlof, C.~W., et al.\ 2004, \aj, 127, 2436 
\bibitem[Yang et al.(2017)]{2017arXiv171111491Y} Yang M., et al., 2017, arXiv:1711.11491 
\bibitem[Yip et al.(2004)]{Yip2004} Yip, C.~W., Connolly, A.~J., Szalay, A.~S., et al.\ 2004, \aj, 128, 585
\bibitem[Yoachim et al.(2009)]{2009AJ....137.4697Y} Yoachim P., McCommas L.~P., Dalcanton J.~J., Williams B.~F., 2009, AJ, 137, 4697 
\bibitem[York et al.(2000)]{York2000} York, D.~G., Adelman, J., Anderson, J.~E., Jr., et al.\ 2000, \aj, 120, 1579
\bibitem[Zhang et al.(2016)]{Zhang2016} Zhang, M., Bakos, G.~{\'A}., Penev, K., et al.\ 2016, \pasp, 128, 035001
\end{thebibliography}

%\begin{thebibliography}{99}
%\end{thebibliography}

%%%%%%%%%%%%%%%%%%%%%%%%%%%%%%%%%%%%%%%%%%%%%%%%%%

%%%%%%%%%%%%%%%%% APPENDICES %%%%%%%%%%%%%%%%%%%%%

%\appendix


%If you want to present additional material which would interrupt the flow of
%% the main paper, it can be placed in an Appendix which appears after the list
% of references.

%%%%%%%%%%%%%%%%%%%%%%%%%%%%%%%%%%%%%%%%%%%%%%%%%%

% Don't change these lines
\bsp	% typesetting comment
\newpage

\appendix
\section{Disk and Stream Fields} 
\label{sec:appendix}

\begin{figure*}
        \includegraphics[width=1.8\columnwidth]{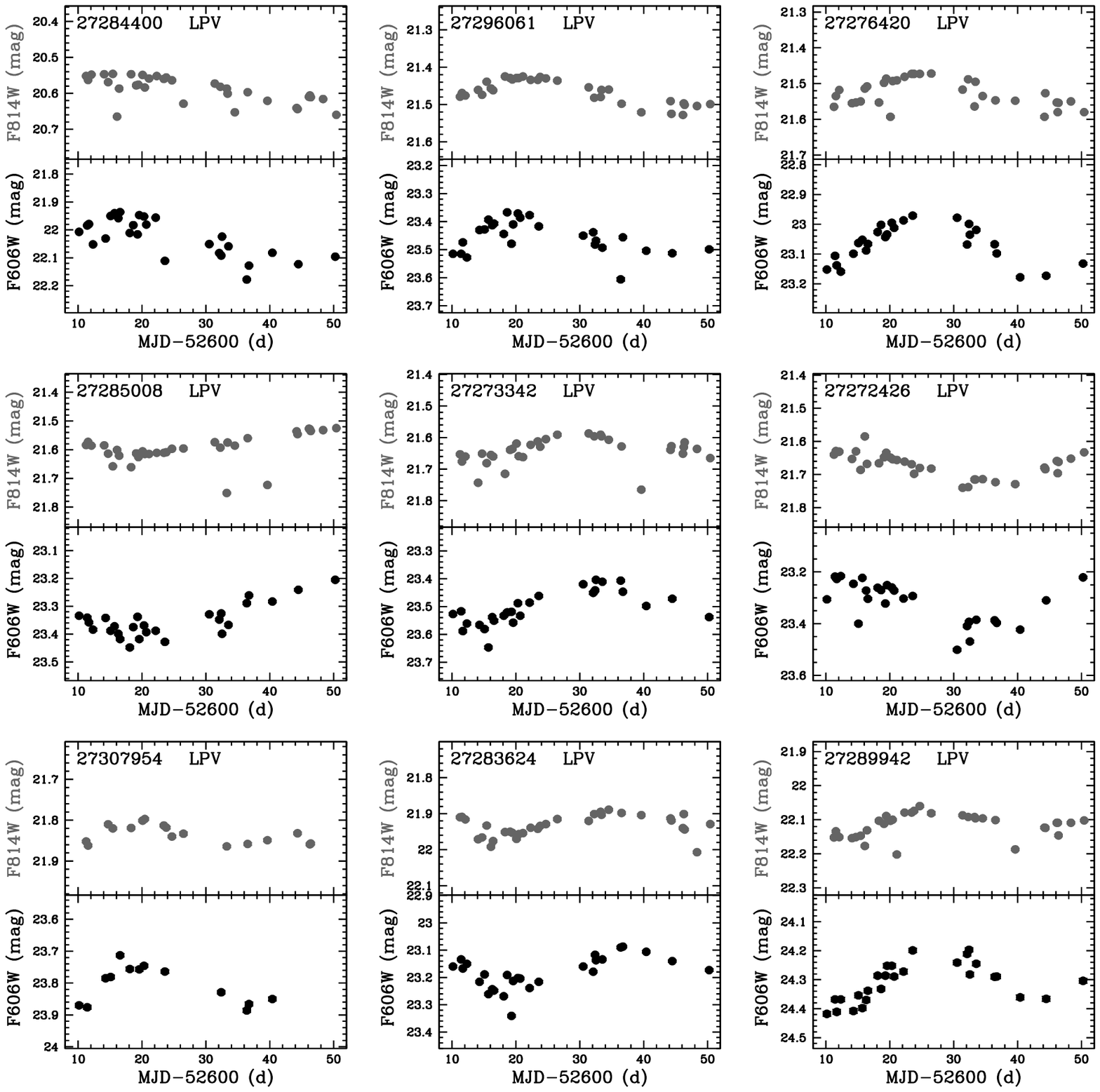}
     \caption{Light curves for the new variables in the Halo11 
field, ordered by F814W magnitude.}
     \label{fig:LCnewConfirmedHalo11}
\end{figure*}

\begin{figure*}
        \includegraphics[width=1.8\columnwidth]{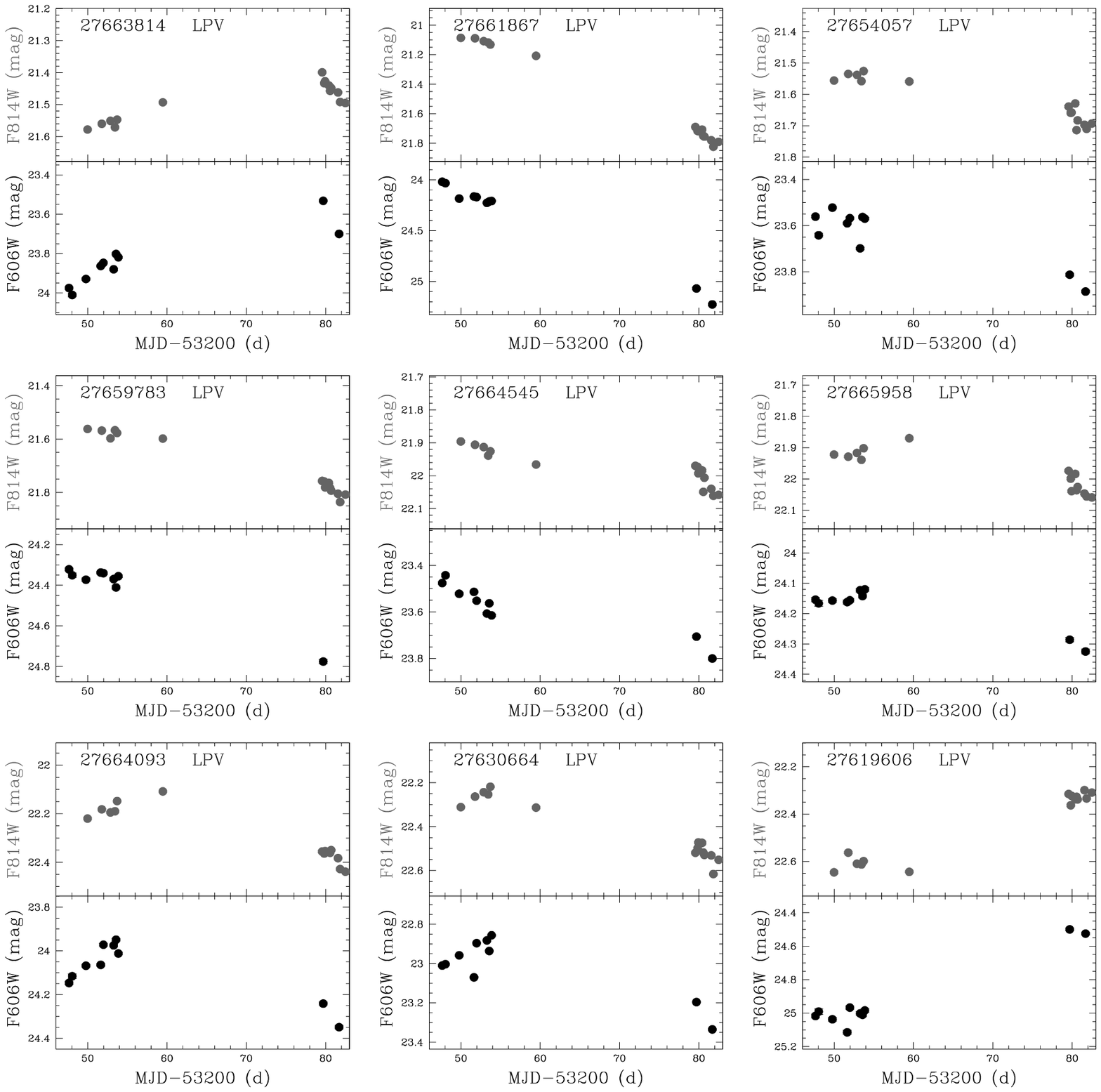}
     \caption{Light curves for the new variables in the Stream 
field, ordered by F814W magnitude.}
     \label{fig:LCnewConfirmedStream}
\end{figure*}

In this appendix we show similar figures as those included in the main
part of the paper but for the Disk and Stream fields. The presentation
and discussion provided for field Halo11 are also applicable to the Disk
and Stream fields unless specifically indicated.
 We also present the LCs of all new variables identified in
the Halo11 (Fig.~\ref{fig:LCnewConfirmedHalo11}), Disk (Fig.~\ref{fig:LCnewConfirmedDisk})
and Stream (Fig.~\ref{fig:LCnewConfirmedStream}) fields.

\begin{figure*}
\begin{centering}
\includegraphics[width=0.48\textwidth,clip=true,trim=1.3cm 0cm 0.7cm 0cm]{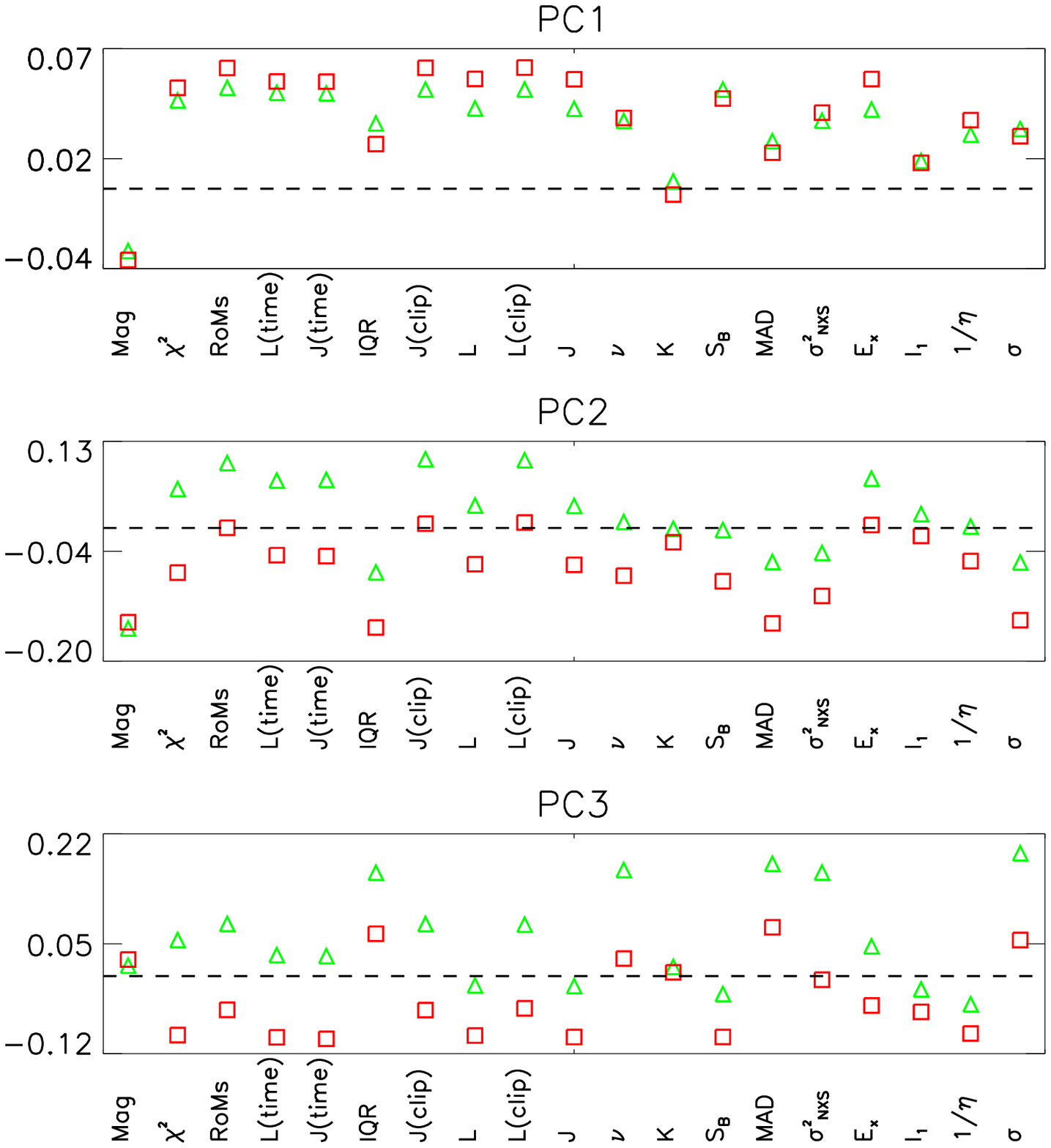}       
\includegraphics[width=0.48\textwidth,clip=true,trim=1.3cm 0cm 0.7cm 0cm]{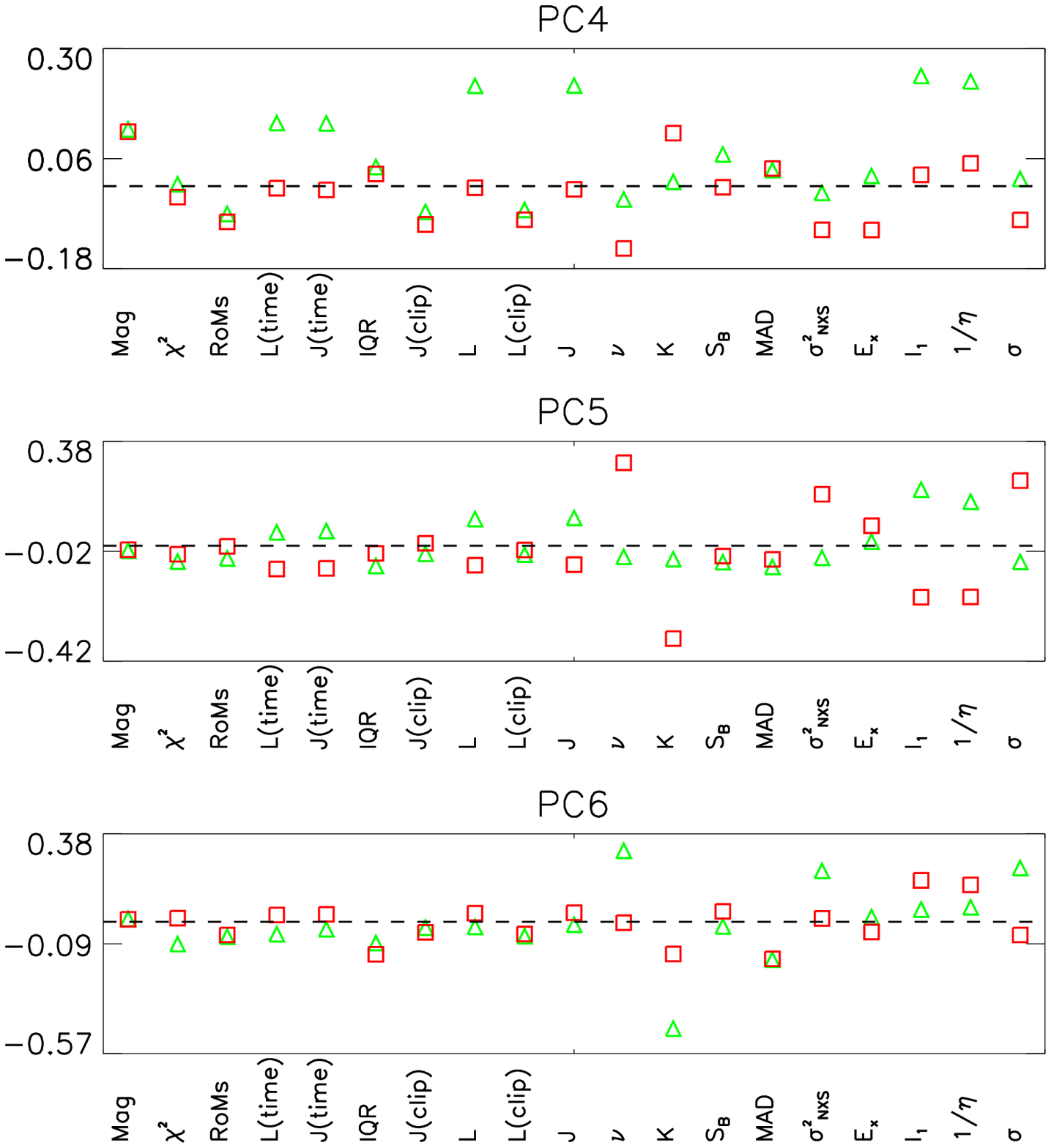}       
    \caption{The first six principal components for the M\,31 Disk data set. The
    dashed line indicates zero contribution of an index to the PC. Green triangles
    correspond to F606W, red squares to F814W.}
\label{fig:disk_pcs_both}
\end{centering}
\end{figure*}

\begin{figure*}
\begin{centering}
\includegraphics[width=0.48\textwidth,clip=true,trim=1.3cm 0cm 0.7cm 0cm]{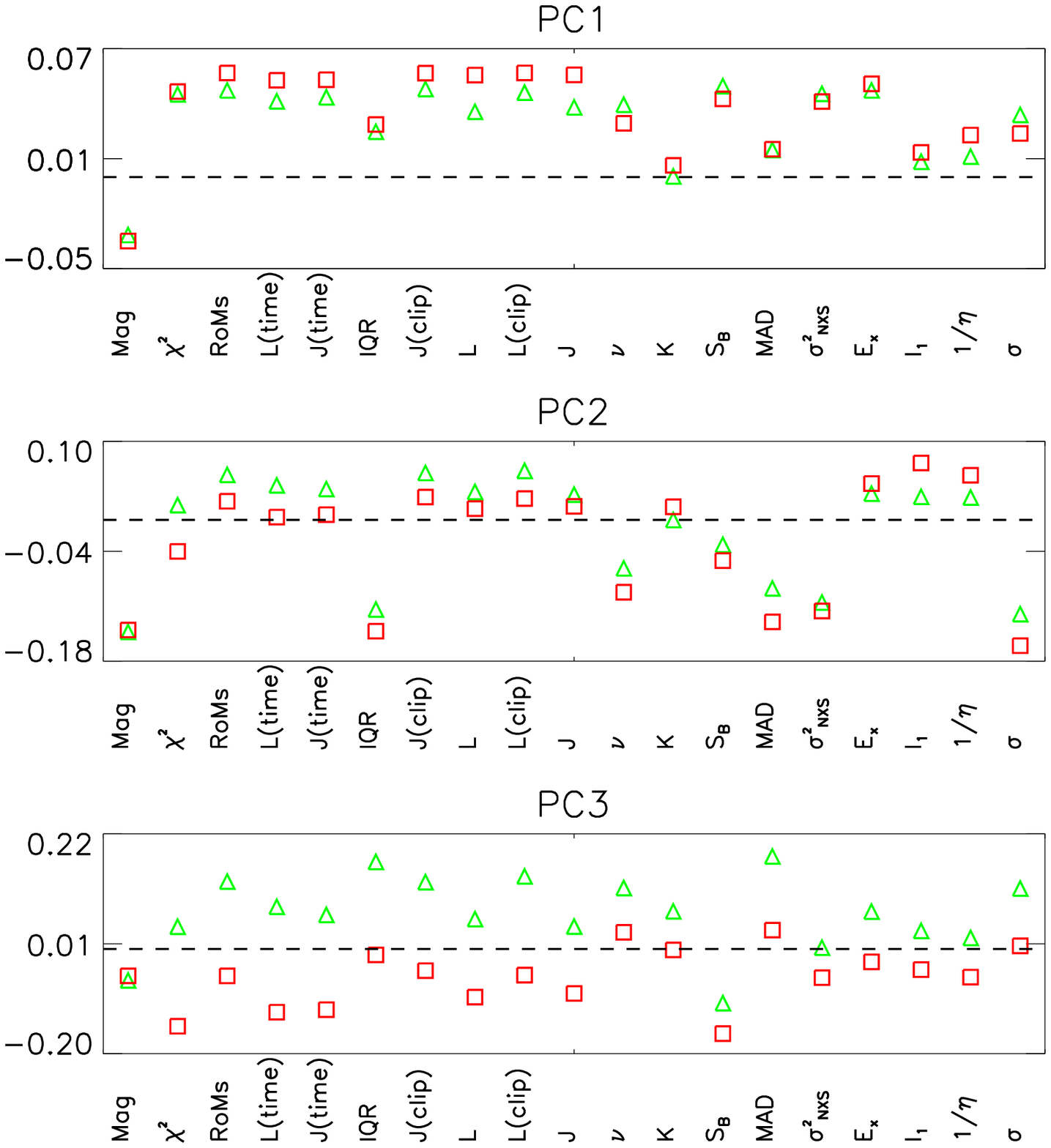}
\includegraphics[width=0.48\textwidth,clip=true,trim=1.3cm 0cm 0.7cm 0cm]{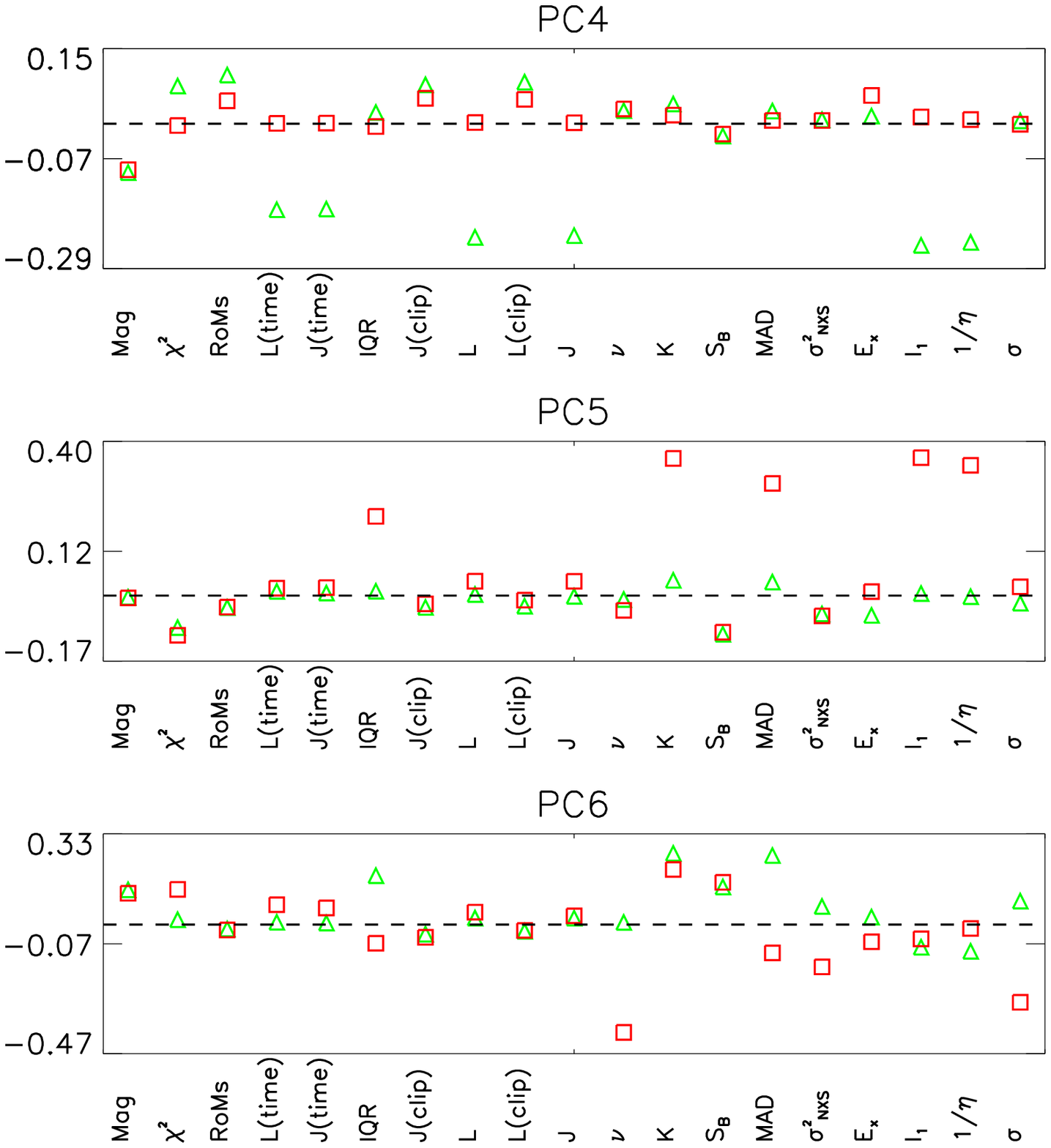}
    \caption{Same as Fig.~\ref{fig:disk_pcs_both} but for the Stream field.}
    \label{fig:stream_pcs_both}
\end{centering}
\end{figure*}

\begin{figure*}
	\includegraphics[width=.85\columnwidth]{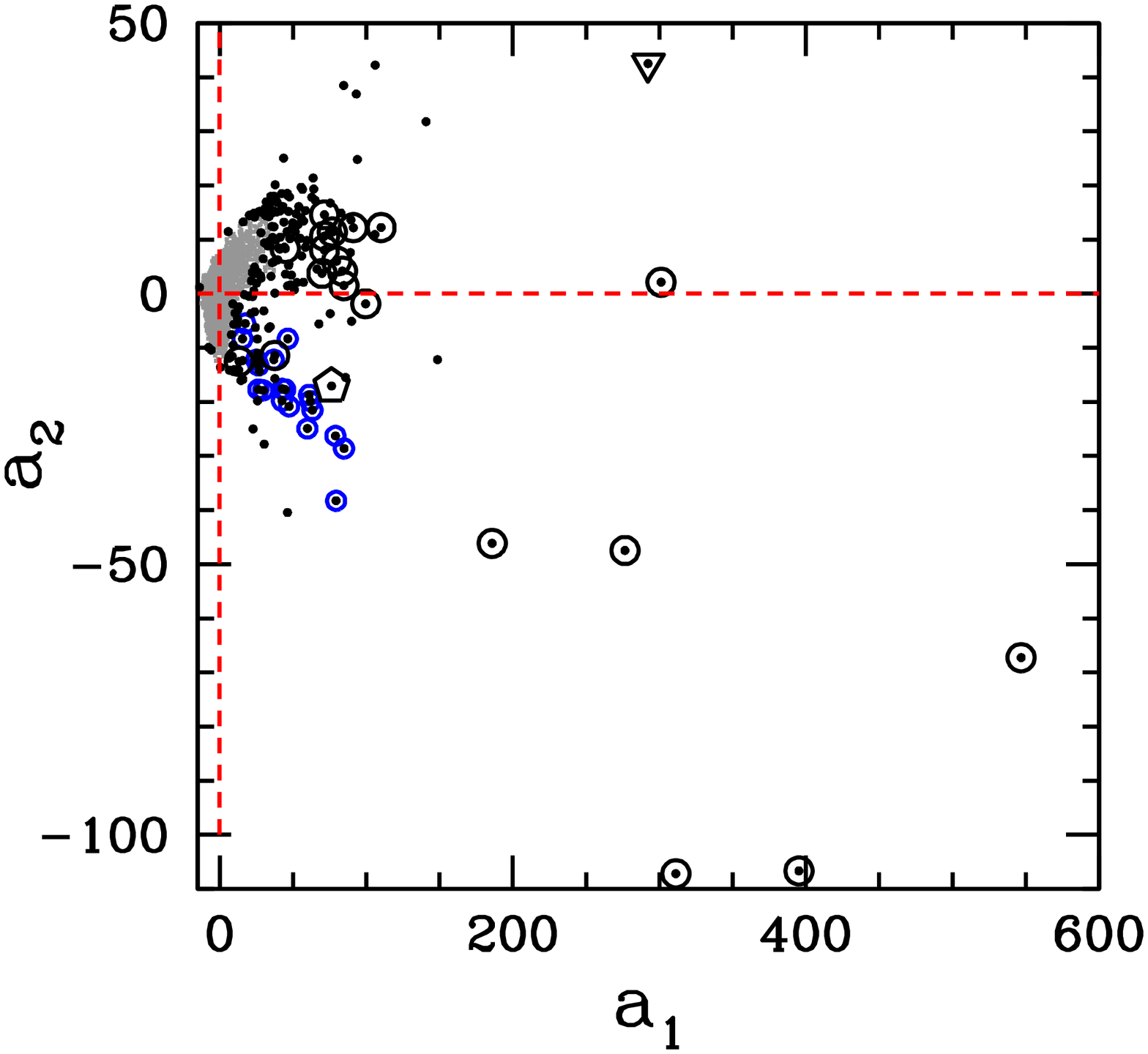}~~~~
	\includegraphics[width=.85\columnwidth]{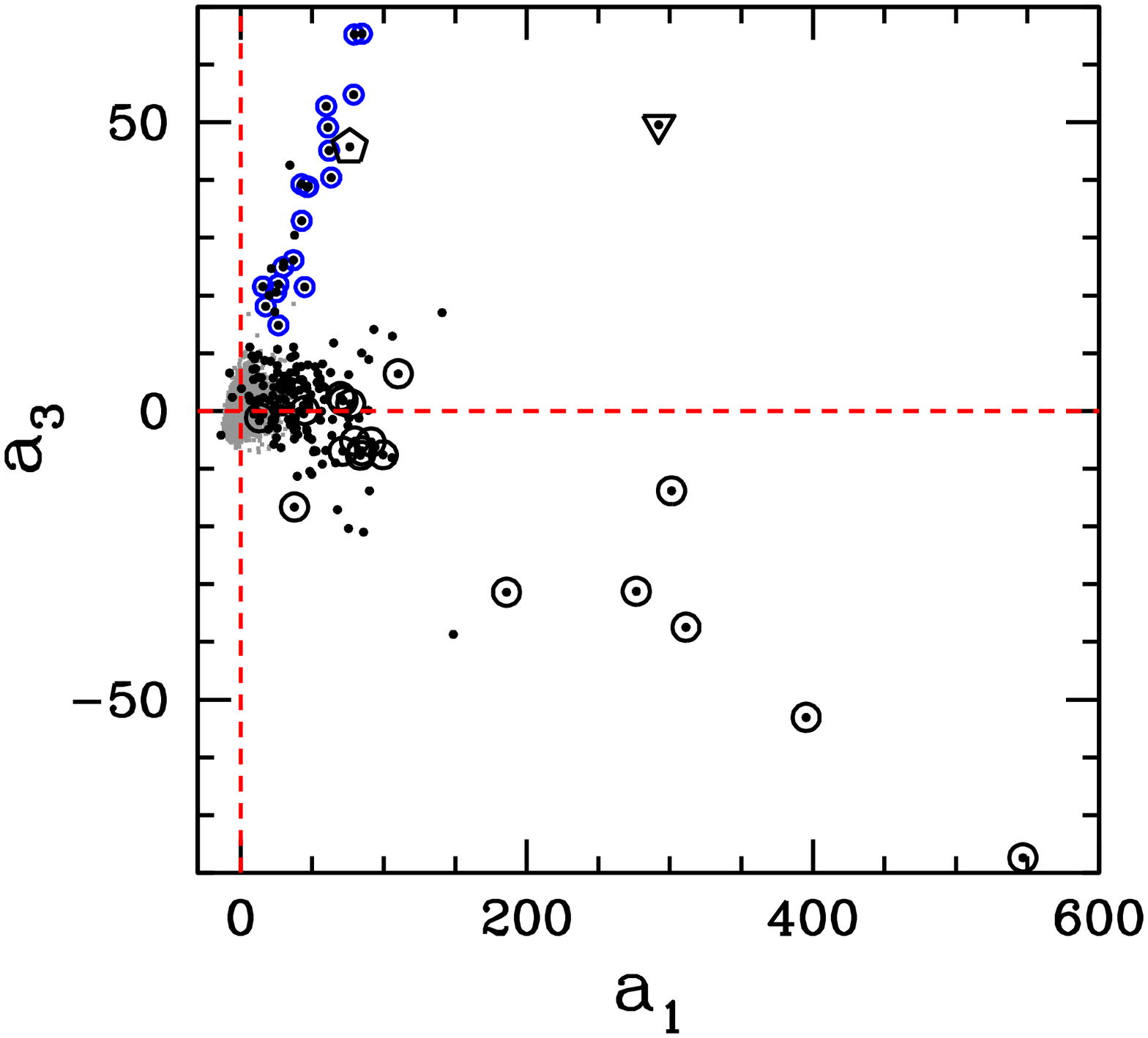}
        \caption{Same as Figure~\ref{fig:halo_admixture}, but for the
          M\,31 Disk. The candidate variables obtained with the method
          described in Section~\ref{sec:selection} are highlighted
          with black points, while newly discovered variables are marked
          with big black open circles. Small blue open circles, the
          black open pentagon and reversed black open triangle highlight
          RR~Lyrae stars, anomalous and classical Cepheid variables,
          respectively, from  \protect\cite{Jeffery2011}. 
}
    \label{fig:disk_admixture}
\end{figure*}

\begin{figure*}
	\includegraphics[width=.85\columnwidth]{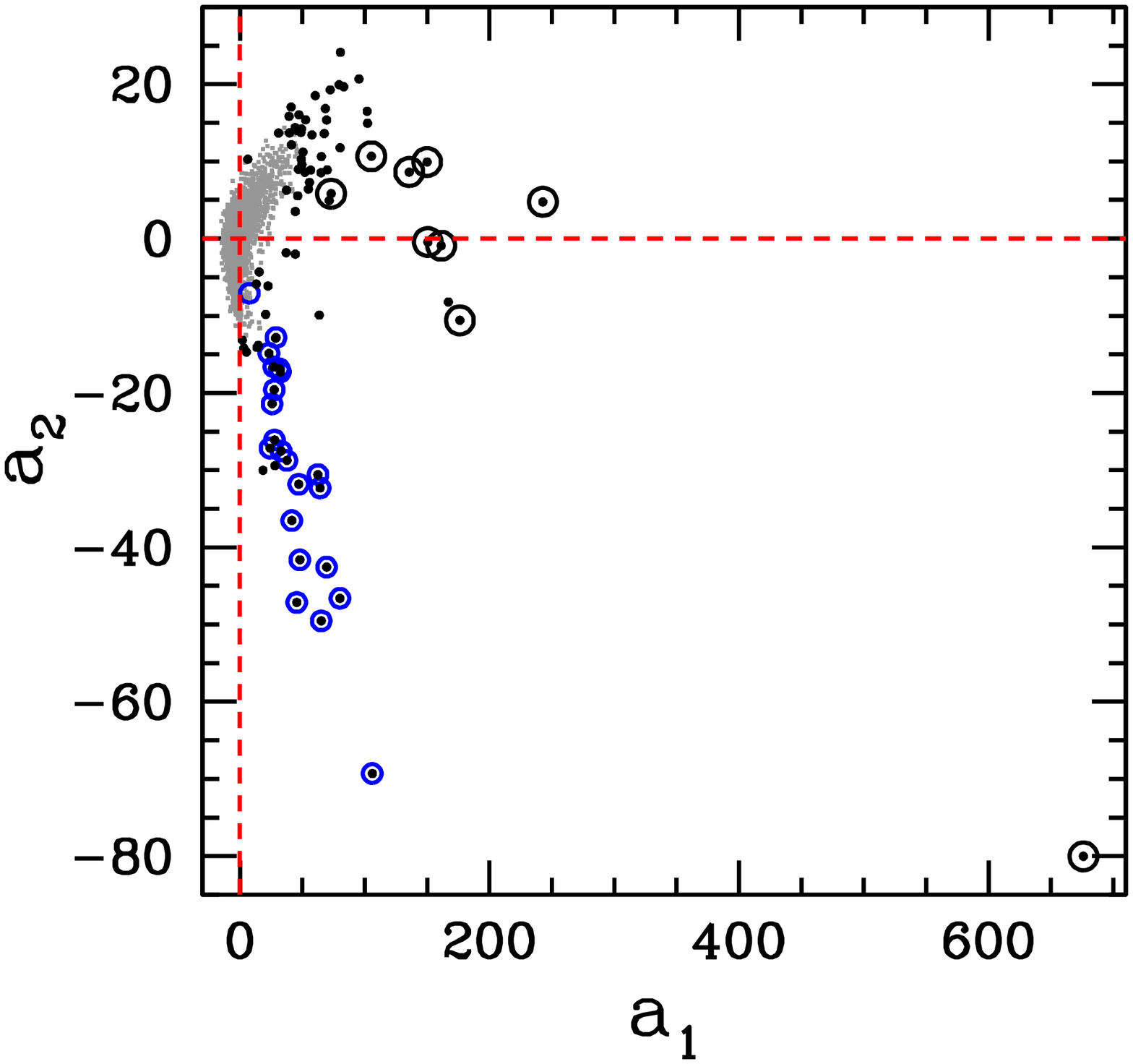}~~~~
	\includegraphics[width=.85\columnwidth]{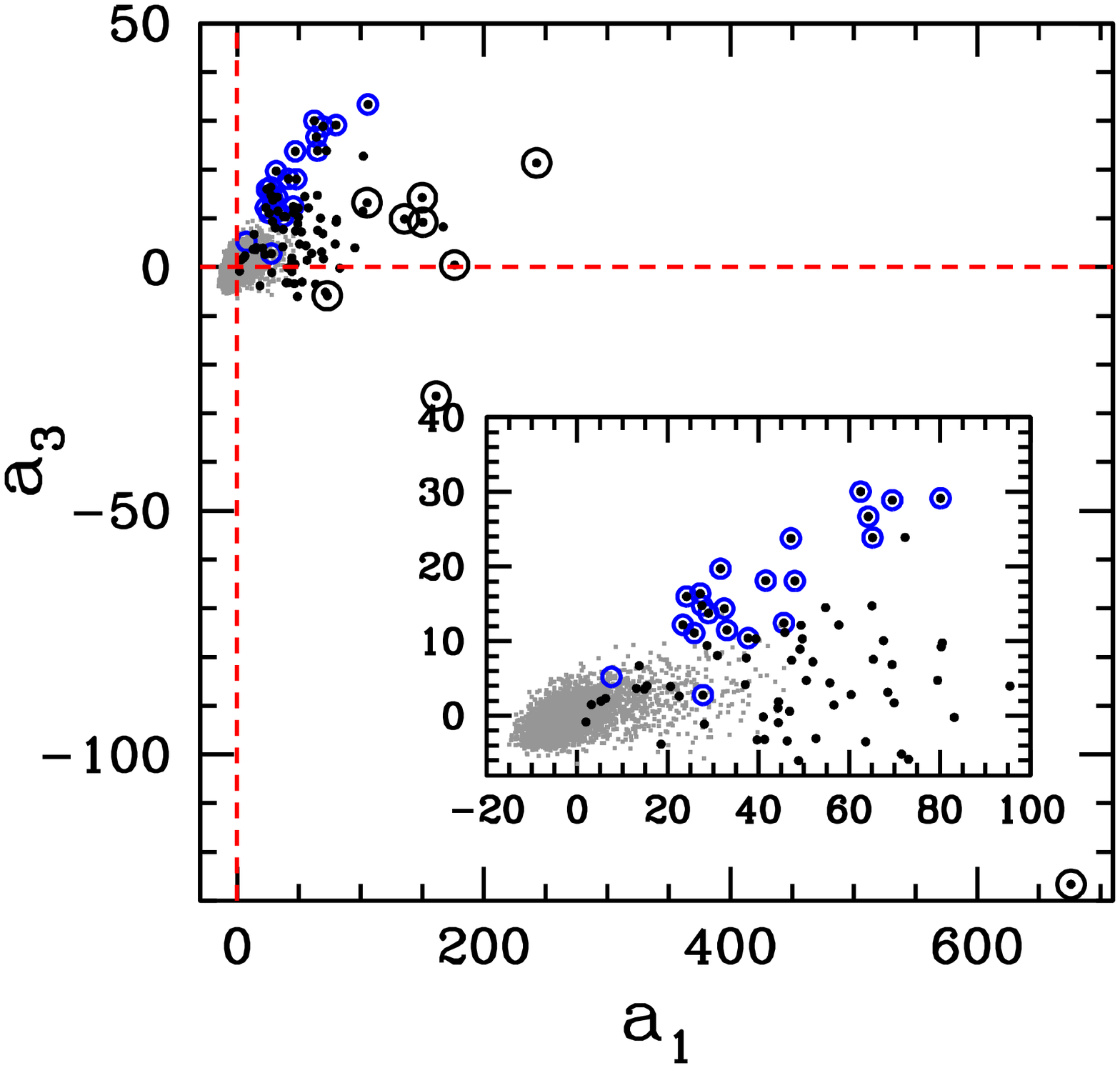}
    \caption{Same as Figure~\ref{fig:halo_admixture}, but for the M\,31
      Stream. The candidate variables obtained with the method described
      in Section~\ref{sec:selection} are highlighted with black
      points, while newly discovered variables are marked with big black
      open circles. Small blue open circles highlight RR~Lyrae variables
      from  \protect\cite{Jeffery2011}. 
}
    \label{fig:stream_admixture}
\end{figure*}

\begin{figure*}
        \includegraphics[width=1.8\columnwidth]{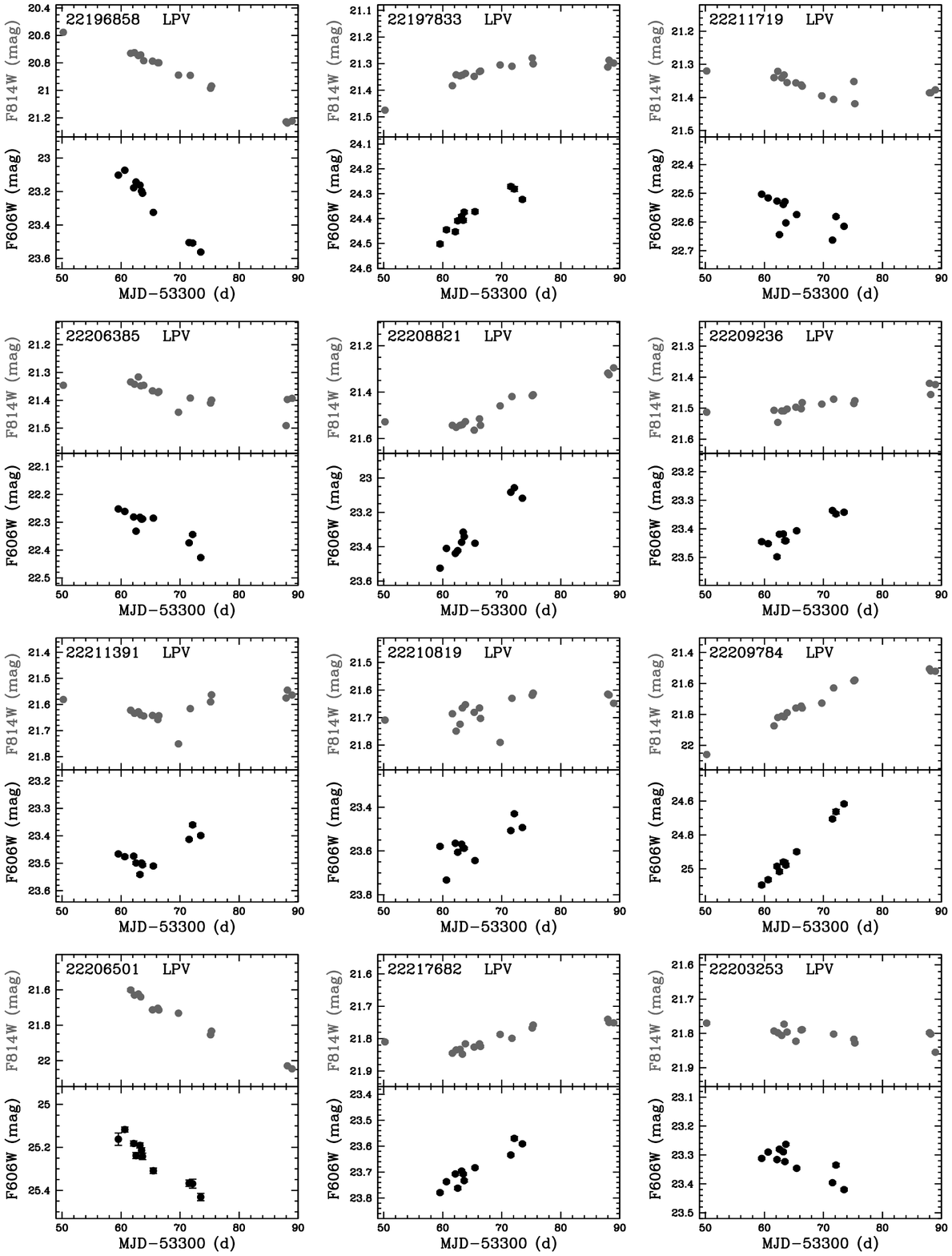} 
     \caption{Light curves for the new variables in the Disk 
field, ordered by F814W magnitude.}
     \label{fig:LCnewConfirmedDisk}
\end{figure*}

\begin{figure*}
\includegraphics[width=1.8\columnwidth]{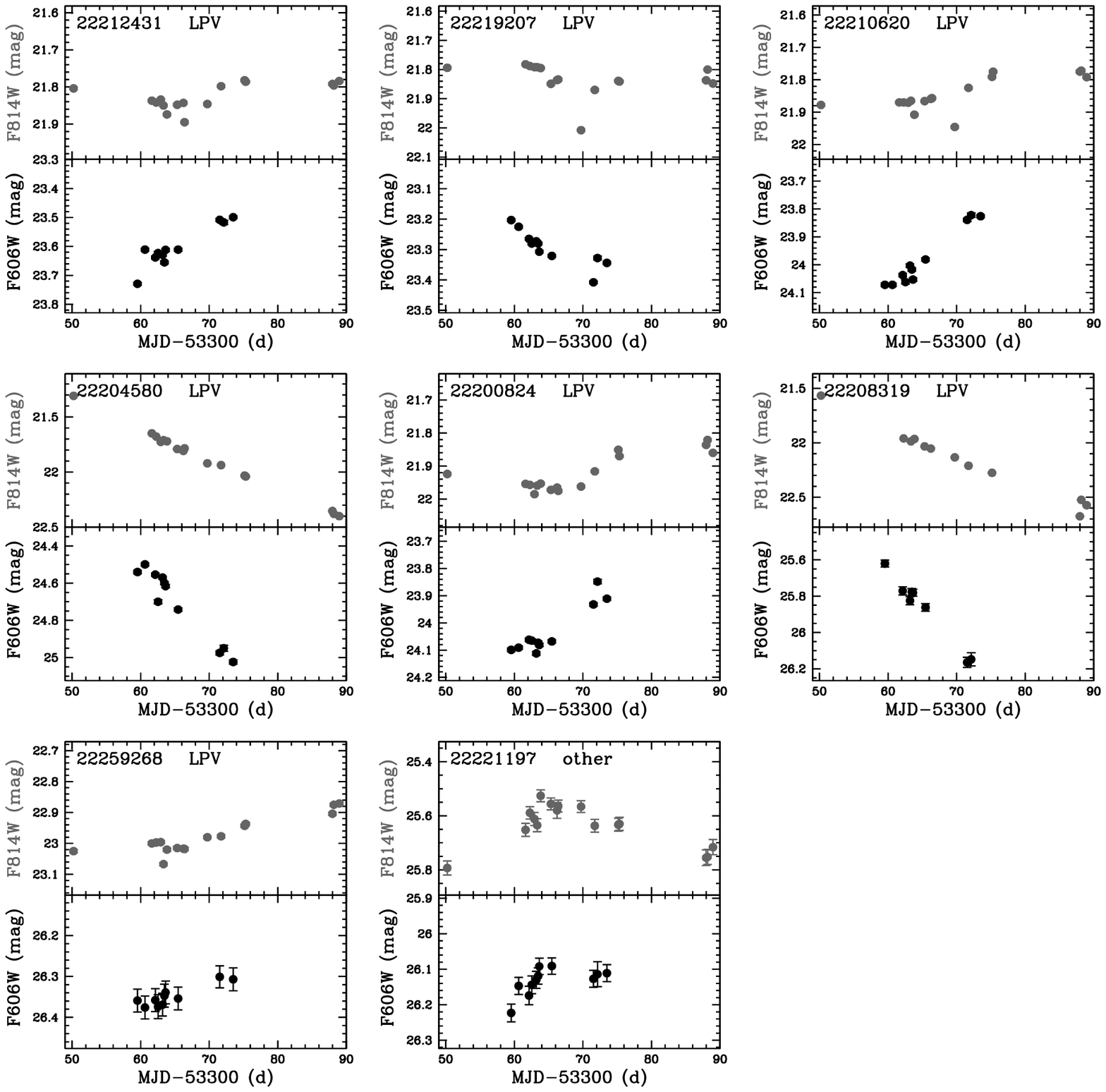} 
\contcaption{}
\end{figure*}

\end{document}